\documentclass[twocolumn,showkeys,amsmath,amssymb,floatfix]{revtex4}
\usepackage{epsfig}
\usepackage{graphicx}
\usepackage{dcolumn}
\usepackage{bm}
\usepackage{amsmath}
\usepackage{amsfonts}
\usepackage{amssymb}

\begin{document}

\title{NMR measurements of hyperpolarized $^{3}\mathrm{He}$ gas diffusion\\ in
high porosity silica aerogels}

\author{Genevi\`{e}ve Tastevin}
 \email{genevieve.tastevin@lkb.ens.fr}
\author{Pierre-Jean Nacher}%
 \email{pierre-jean.nacher@lkb.ens.fr}
\affiliation{Laboratoire Kastler Brossel, 24 rue Lhomond, 75231 Paris Cedex 05, France}
  
\thanks{Research unit of the Ecole Normale Superieure, and of the Pierre et Marie Curie
University (Paris 6), associated to the Centre
National de la Recherche Scientifique (UMR 8552).}

\date{\today}

\begin{abstract}
Hyperpolarized $^{3}\mathrm{He}$ is used to nondestructively probe by NMR the
structure of custom-made and commercial silica aerogels (97\% and 98.5\% porous). Large 
spin-echo signals are obtained at room temperature and very
low magnetic field (2~$\mathrm{mT}$) even with small amounts of gas.
Attenuation induced by applied field gradients results from the combined
effects of gas diffusion and confinement by the porous medium on atomic
motion. Nitrogen is used as a buffer gas to reach equivalent $^{3}\mathrm{He}$
pressures ranging from 5~$\mathrm{mbars}$ to 3.5~$\mathrm{bars}$. The observed
pressure dependence suggests a non-uniform structure of the aerogels on length
scales up to tens of micrometers. A description by broad phenomenological
distributions of mean free paths is proposed, and quantitatively discussed by
comparison to numerical calculations. The investigated aerogel samples exhibit
different effective diffusion characteristics despite comparable nominal porosities.
\end{abstract}



\maketitle

Article published in the Journal of Chemical Physics. 

Reference : J. Chem. Phys. \textbf{123}, 064506 (2005)

\section{Introduction}

Silica aerogels are highly porous solids obtained by sol-gel process
\cite{Fricke97}. Their very open structure consists of a delicate, highly
ramified backbone network of randomly interconnected thin silica strands,
formed by aggregation of small silica clusters \cite{Schaefer86,Fricke88}.
Mechanically robust, they exhibit very low densities, thermal conductivities,
dielectric constants and sound velocities, as well as poor chemical reactivity
and large specific areas. Despite high synthesizing costs and difficulties,
they have been commercially produced for civil and military use and they have
found applications in thermal insulation, manufacture of materials and
coatings, catalysis, gas storage and filtering, Cerenkov radiation detection,
etc. In science, their unique properties make them attractive for cutting-edge
studies in optics, acoustics, chemical physics, low-temperature physics,
nuclear physics, space investigations, laser tests, microelectronics,
electrical engineering, life sciences and bioengineering
\cite{Fricke97,Pierre02,Akimov03}. The density and microstructure of the
aerogels can be precisely controlled by adjusting the growth chemistry
\cite{Pierre02,Ferri91,Schaefer89,Vacher88}. This makes them particularly
suited for a variety of investigations on bulk, film or layer systems in
weakly confined random geometries.

Considerable effort has been devoted to the study of phase transitions in
aerogel \cite{Frisken91,Reppy92,Wong93, Porto99}, in particular with helium at
low temperature\ to probe statistical and quantum effects
\cite{Reppy92,Porto99,Chan96}. The silica network provides a structural
disorder background\ to the fluid where it is imbedded, but occupies only a
very small fraction of the sample volume. Recently, extremely high (up to
99.5\%) porosity aerogels have been used to study impurity scattering effects
on unconventional pairing states in superfluid $^{3}\mathrm{He}$
\cite{Lawes00,Hanninen03} and changes of critical exponents in superfluid
$^{4}\mathrm{He}$ \cite{Yoon98}. Strong modifications of the phase diagram
occur when the correlation length of the order parameter in the fluid becomes
comparable to that of aerogel. The microstructure of the samples used in the
experiments must hence be characterized at the appropriate length scales, that
range from nanometers (15$-$80~$\mathrm{nm}$ in superfluid $^{3}\mathrm{He}$
\cite{Porto96}, 3$-$100~$\mathrm{nm}$ in superfluid $^{4}\mathrm{He}$
\cite{Singsaas84,Yoon98}) to tens of microns (mean free path, mfp, of
unpaired $^{3}\mathrm{He}$ quasiparticles at low pressure and temperature
\cite{Porto99}).

In contrast with Vycor glass or porous gold, for instance, there is no single
or dominant length scale to characterize the distribution of pores and silica
in aerogel. Aerogel is known to be fractal at very small scales, a feature
revealed both by small angle scattering (SAS) measurements with x rays or
neutrons, and computer simulations of the growth process
\cite{Schaefer86,Boukenter89,Hasmy94}. Self-similar structures range over
correlation length scales from a few nanometers to hundreds of nanometers
(depending on catalysis conditions and kinetics of the gelation process),
above which the medium appears to be fairly homogeneous. Actual homogeneity at
microscopic level, and relevance of a description solely based on the fractal
dimension and one correlation length parameter, are\ key issues for the
interpretation of all physical measurements \cite{Porto99}, acutely debated in
the case of superfluid $^{3}\mathrm{He}$ and $^{4}\mathrm{He}$ in aerogel
\cite{Hanninen03,Vasquez03}. Discrepancies between experimental results are
common and, most generally, tentatively attributed to unknown sample-to-sample
microscopic variations (due to growth conditions, or insufficient
characterization \cite{Porto99}). In addition, meso- or macroscopic
inhomogeneities may develop during sample preparation, mounting, or use at low
temperature (due to mechanical or thermal stress \cite{Eska03}).
Unfortunately, conventional techniques such as mercury intrusion porometry,
thermoporometry, or nitrogen adsorption/desorption, are not suitable to
characterize the aerogel pore structure accurately \cite{Scherer98}.

Nuclear magnetic resonance (NMR) is an established noninvasive technique to
characterize porous media on a microscopic\ scale \cite{Mitra97,Callaghan}.
Gas probes are of particular interest, because they can easily penetrate the
porous matrix and exhibit fast diffusion rates that are conveniently varied
over a wide range of values by pressure control. Compared to other inert
species, $^{3}\mathrm{He}$ may be considered as an ideal candidate probe for
diffusion measurements: this spin-1/2 atom has long intrinsic relaxation
times, and a resonance frequency fairly insensitive to most changes of its
local environment \cite{Mazitov93,Seydoux96}. Suitable sensitivity is usually
achieved using both high pressure and high magnetic field
\cite{Seydoux96,Lizak91} to compensate for the low density of the gas. With
laser optical pumping techniques, high nuclear polarizations
(hyperpolarizations) can be achieved independently of the magnetic field
strength, and very high magnetization densities are obtained even at low
pressures. Submillimetric imaging and diffusion measurements of
hyperpolarized gases in free or confined spaces have been demonstrated both at
high and low magnetic field \cite{Raftery91,Song95,Saam96,Tseng98}, and used
for various purposes including \textit{in vivo} characterization of the lung alveolar
airspaces for medical applications
\cite{Salerno01,vanBeek04,Durand02,Bidinosti03}. In the present work, we use
hyperpolarized $^{3}\mathrm{He}$ NMR to non-destructively probe the void
spaces in high porosity silica aerogels identical to those used in most low-temperature 
experiments. With probe gas mfp's ranging from a fraction of
micrometer to a few tens of micrometers when pressure is varied, this
technique yields a unique type of information on the structure of these materials.

We had previously performed preliminary measurements on aerogel samples to test
the feasibility of room temperature NMR investigations \cite{GT00,GT00b,GG01}. They suggested
a non-uniform distribution of density of silica strands in this system, in
contrast, for instance, with the single quasiparticle mfp value obtained
by spin diffusion measurements in $^{3}\mathrm{He}$-$^{4}\mathrm{He}$ dilute
solutions \cite{Candela98}. We have now undertaken a more extended investigation
to confirm this result, using an improved experimental setup. In this
article, we report on a systematic study over a broader range of gas pressures
that has been carried out to probe several samples of custom-made and
commercially produced aerogels. Our former conclusions \cite{GT00b,GG01} are
confirmed. We show that all aerogels exhibit the same qualitative behavior,
characterized by a pressure dependence of the gas diffusion coefficient that
excludes a description of the aerogel by a single characteristic length.
Significant differences are observed between the various samples, that do not
correlate with the specified aerogel porosities and may rather be associated
to differences in growth conditions. We also propose a phenomenological
description of gas diffusion in aerogel based on broad power-law distributions
of mfp's. Their main features are quantitatively derived from the
NMR data analysis, and discussed on the basis of computer calculations.

\section{Experiment}

\subsection{Aerogel samples}

Six samples of aerogel with 97\% and 98.5\% nominal porosities have been studied.
One is custom made, and all others come from specialized manufacturers.

Sample M is a 98.5\% open custom-made aerogel prepared by hydrolysis and
polymerization of tetramethoxysilane, dilution in acetonitrile, gelation after
addition of ammonia, and supercritical drying in an excess of solvent to
preserve the delicate structure \cite{Tillotson88}. Because of the base
catalyzing, the gelation process is believed to be diffusion-limited
aggregation of small silica clusters (1$-$3~$\mathrm{nm}$ in diameter). This
type of aerogel, used in many low-temperature experiments, has been fairly
characterized at small length scales using experimental (scanning and
transmission electron micrography, SEM\ and TEM; SAS of x rays and neutrons)
and numerical (diffusion-limited cluster aggregation, DLCA, growth models)
techniques \cite{Hasmy94}. The nanostructure of computer-generated aerogels,
nicely reproducing that of real samples probed by SEM/TEM and SAS, can be most
conveniently visualized (see, e.g., \cite{Porto99,Hasmy94} for 2D or 3D
images) and statistically analyzed. On custom-made 98.5\% samples similar to
ours, SAS of x rays shows a fractal-like distribution of silica extending up
to a length scale of 65$-$100~$\mathrm{nm}$ \cite{Chan96,Yoon98}, from which a
purely geometrical mfp on the order of 270~$\mathrm{nm}$ can be calculated
\cite{PortoPhD}. In a comparable DLCA model (98.2\% porosity), the very dilute
aerogel structure is found to be organized so that half of the widely open
volume lies farther than 10~$\mathrm{nm}$ from some silica, while none of it
is farther than 35~$\mathrm{nm}$ \cite{PortoPhD}. Recently, path-length
distributions have been analyzed in simulated aerogels using randomly sampling
straight-line trajectories \cite{Haard00}, and finite-size hard spheres
displacements to mimic experimental conditions \cite{McElroy03}. The estimated
$^{3}\mathrm{He}$ quasiparticle mfp significantly exceeds in aerogel that
of a randomly distributed medium of identical volume density (180 and
120~$\mathrm{nm}$, respectively, for a 98.5\% open sample \cite{McElroy03}).

Aerogel M has been grown in a cylindrical glass container (inner diameter
12~$\mathrm{mm}$, height 37~$\mathrm{mm}$) partly filled with a set of three
glass tubes (diameter 6$\times$4~$\mathrm{mm}$, height 30~$\mathrm{mm}$, and axes
parallel to that of the container). It has been produced as a transparent,
crack-free sample attached to the glass walls. An epoxy end cap for the
aerogel container is glued to allow leak-tight connection to the $^{3}$He
polarization system, with negligible dead volume between the free aerogel
surface and the epoxy cap (Stycast 1266 A/B, Emerson and Cuming, Billerica, MA, USA).

Sample A is a 97\% porous aerogel with nominal specific area
465~$\mathrm{m}^{2}\mathrm{/g}$ manufactured by Airglass, Staffanstrop,
Sweden. The cylindrical sample (diameter 12.5~$\mathrm{mm}$, height
10.8~$\mathrm{mm}$) is enclosed in a matching PMMA \cite{PMMA} container.
The container and end cap are machined so that the aerogel gently fits in,
avoiding mechanical stress ($\sim$0.05~$\mathrm{mm}$ gap size).

Samples J$_{1}-$J$_{4}$ are four pieces of 97\% porous aerogel with
nominal specific area 510~$\mathrm{m}^{2}\mathrm{/g}$ manufactured by
Matsushita Electric Works, Osaka, Japan. J$_{1}$ and J$_{2}$ originally had
identical parallelepipedic shapes (sections 15$\times$15.5~$\mathrm{mm}%
^{2}\mathrm{,}$ thicknesses 9.7~$\mathrm{mm)}$. J$_{1}$, accidentally crushed
before use, is reduced in thickness by 1.7$-$2.7~$\mathrm{mm}$ (from end to
end along its longest dimension, with a noticeable transverse fracture line).
J$_{3}$ and J$_{4}$ have cylindrical shapes (diameters 9.1 and
7.7~$\mathrm{mm}$, lengths 10.7 and 10.5~$\mathrm{mm}$,
respectively). Their original rough ends have been cut flat using a silicon
carbide disk driven at very high speed (65~000~rpm). Viewed through these new
flat ends, the aerogels are perfectly transparent and display no macroscopic
flaw. All four samples are enclosed in tightly fitting PMMA containers (apart
from a 0$-$1~$\mathrm{mm}$ gap left open for the trapezoidal sample J$_{1}$).

An empty PMMA container, identical to that of aerogel A (diameter
12.5~$\mathrm{mm}$, height 10.8~$\mathrm{mm}$), is used as a reference cell
for free diffusion measurements.

Polypropylene Luer fittings are glued to the PMMA containers to allow
connection to the $^{3}\mathrm{He}$ polarization system,\ through a
10-$\mathrm{cm}$-long Tygon tube (Masterflex, Cole-Parmer Instrument
Co.,Vernon Hills IL, USA) with 1.6-$\mathrm{mm}$ inner diameter. A 0.7$\times
$1.1-$\mathrm{mm}$ Teflon tube is inserted into the Tygon filling tube to reduce the
dead volume. This also increases the gas flow impedance, hence limits the
pressure change rates inside the aerogel sample.

There has been no attempt to clean the aerogels. The surface may have retained
traces of preparation chemicals, as well as $\mathrm{O}_{2}$ or moisture
adsorbed during storage. During experiments, the samples are under vacuum, or
in contact with high purity $^{3}\mathrm{He}$ or $\mathrm{N}_{2}$ gas. A
totally negligible amount of\ He atoms may dissolve in the silica
\cite{Seydoux96}. The aerogel behavior is controlled during and after use by
visual inspection. Sample M remains clearly well attached to the glass walls.
Several cracks are present in its bulk part, between the free surface and the
top of the inner tubes that presumably provide mechanical support to the
aerogel. They developed during prior use, due to rough filling and emptying
procedures \cite{GT00}, but remained unaffected both by the breaking of
one inner tube and the present series of measurements. All off-the-shelf
samples (A, J$_{1}-$J$_{4}$) are used for the first time. Through the PMMA
side walls, J$_{1}$ and J$_{2}$ can be observed to expand and shrink slightly
as pressure varies from 0 to 1~$\mathrm{atm}$. Very few cracks developed
inside these commercial aerogels, allowed to move freely inside the containers
by an irreversible decrease of all their dimensions (by about 5\%$-$10\%) observed
after a few pressure cycles.

\subsection{NMR setup}

The pulsed NMR experiments are performed at low frequency (62.4~$\mathrm{kHz}
$) using a homemade system, in the homogeneous static field used for 
optical pumping (OP) described in Section~\ref{section2.3}. We use free induction
decay (FID) signals to
monitor the $^{3}\mathrm{He}$ magnetization, and spin-echoes generated by a
standard Carr-Purcell-Meiboom-Gill (CPMG) sequence with an applied gradient to
measure diffusion \cite{CPMG}.

\begin{figure}[h]
\includegraphics[clip,width=3.15in]{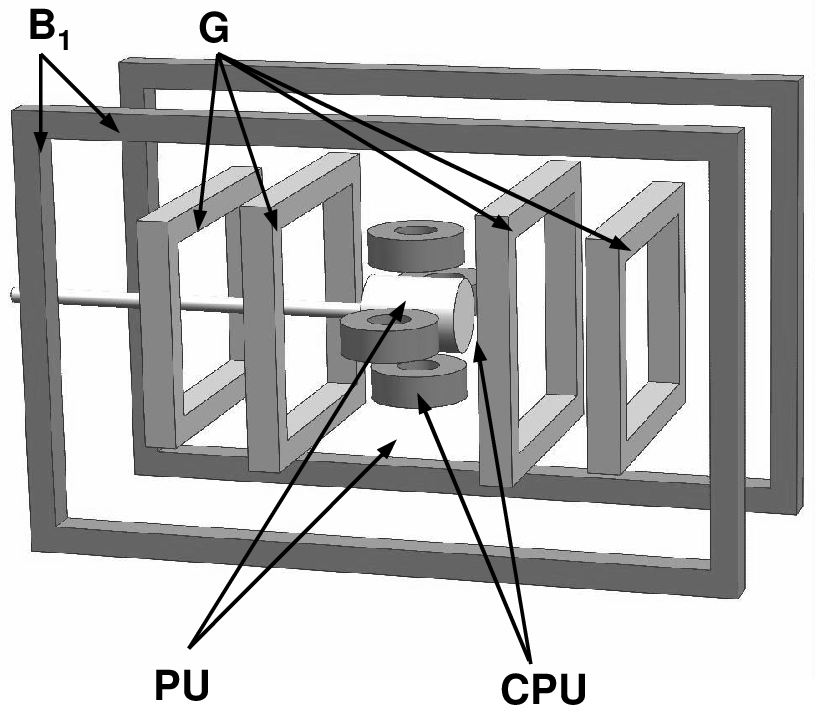}
\caption{Sketch of the NMR coils. G: set of four gradient coils (45$\times$42 (45$\times
$35)-$\mathrm{mm}$ size, 4-$\mathrm{mm}$ horizontal width, 38
(78)-$\mathrm{mm}$ average spacing, and 9 (25) turns each, for inner (outer)
coils, respectively; 0.75-$\mathrm{mm}$ diam. wire). B$_{1}$: oversized
induction coils (75$\times$125-$\mathrm{mm}$ size, 4-$\mathrm{mm}$ horizontal
width, 52.5-$\mathrm{mm}$ average spacing, 25 turns each; 0.4-$\mathrm{mm}$
diam. wire). PU: pickup coils, and CPU: extra detection coils for rf noise
compensation - see text - (10$\times$18-$\mathrm{mm}$ diam., 6-$\mathrm{mm}$
width, 440 turns each; 0.2-$\mathrm{mm}$ diam. wire). A thin filling tube is
attached to the aerogel container (here depicted with cylindrical shape).}
\label{fig1}
\end{figure}

All NMR and gradient coils depicted in Fig.~\ref{fig1} are copper wire
windings held by PMMA stands.
They are
surrounded by a closed copper box (23$\times$16$\times$10~$\mathrm{cm}^{3}$,
thickness of 2~$\mathrm{mm}$, not shown) for passive radio-frequency (rf)
shielding. Two untuned, oversized rectangular induction coils are used to
generate a very uniform rf field over the sample. Typically, a 90$%
{{}^\circ}%
$ tip angle is achieved by a 0.5-$\mathrm{ms}$ rf pulse with 11.5-$\mathrm{V}%
\ $rms excitation. 180$%
{{}^\circ}%
$ pulses induce negligible magnetization losses, less than 10\% after 1000
echoes refocused without applied gradient. Two circular detection coils lie in
conventional Helmholtz configuration close to the sample (``PU'' coils in
Fig.~\ref{fig1}).\ We use a pair of identical coils (``CPU'' coils in
Fig.~\ref{fig1}), also implemented next to the sample and connected in
opposition with the previous ones, to actively shield the device from pick-up
noise. These two extra coils detect a noise signal from distant rf sources
that almost exactly compensates for that induced in the main detection coils.
Induction and detection coils are oriented so as to minimize the rf
cross-talk, i.e., mutual induction.\ The set of detection coils is tuned to
Larmor frequency with a parallel capacitor, and the quality factor is moderate
($Q$=43). The measured sensitivity corresponds to an expected 90$%
{{}^\circ}%
$ FID signal of 43-$\mathrm{\mu V}$ rms amplitude per $\mathrm{mbar}$ pressure
and $\mathrm{cm}^{3}$ of $^{3}\mathrm{He}$ gas with 100\% nuclear
polarization. Two pairs of rectangular coils, connected in series, are
designed to produce a uniform horizontal magnetic field gradient
(25~$\mathrm{mT/m/A}$) parallel to the static field axis. Computed relative
variations over 3~$\mathrm{cm}$ around the center do not exceed 3\%.

We use a dual-phase lock-in amplifier (LIA; model 7220, PerkinElmer
Instuments, Oak Ridge Tennessee, USA) both to collect the NMR signal and to
generate the resonant rf voltage. Rf pulse gating, dephasing, and
amplification are handled by a personal computer (PC)-controlled home-built circuitry. Gradient
and rf pulse sequencing, as well as data collection via an A/D converter
(sampling frequency: 10~$\mathrm{kHz}$), are managed using a laboratory-written software.

\subsection{$^{3}$He hyperpolarization system\label{section2.3}}

\begin{figure}[h]
\includegraphics[clip,width=3.15 in]{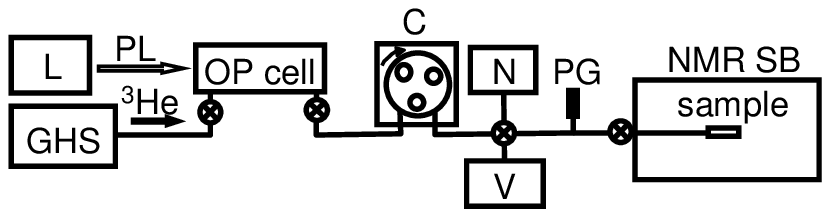} 
\caption{Block diagram of the hyperpolarization system. L: laser source. GHS: gas
handling system. PL: pump laser beam. OP\ cell: optical pumping glass cell. C:
peristaltic compressor. N: Nitrogen gas supply. V: vacuum system. PG: pressure
gauge. NMR\ SB: shielding box surrounding the sample and NMR setup sketched
in Fig.~\ref{fig1}.}
\label{fig2}
\end{figure}

$^{3}\mathrm{He}$ gas is hyperpolarized inhouse using the metastability
exchange optical pumping technique. We use a home-built system that involves
the basic units indicated by the block diagram in Fig.~\ref{fig2}.
The gas
is polarized at a pressure of about 2~$\mathrm{mbars}$ in a cylindrical glass
cell (50-$\mathrm{cm}$ length, 6-$\mathrm{cm}$ diameter). The
1083-$\mathrm{nm}$ light source is a 2-$\mathrm{W}$ ytterbium fiber laser
(model OI-FL1083-2W, Keopsys, Lannion, France). The\ OP beam is expanded and
collimated (2-$\mathrm{cm}$ full width at half maximum), and circularly
polarized. A mirror back-reflects it to enhance light absorption by the small
fraction (typically 1~$\mathrm{ppm}$) of $^{3}\mathrm{He}$ atoms excited to
the metastable state by a weak rf\ discharge (3-$\mathrm{MHz}$ frequency,
2-$\mathrm{W}$ power). A dedicated peristaltic compressor is used to extract
the polarized gas from the OP cell and transfer it into the sample container
\cite{PJN99}. Upstream of the OP cell, the gas handling unit (not shown)
includes a $^{3}\mathrm{He}$ bottle (purity grade higher than 99.99\%), a
getter to keep impurities level well below 1ppm (model GC50, SAES Getters,
Lainate, Italy), a high vacuum system (a turbomolecular pump backed by a
membrane pump, electropolished stainless steel tubes and valves, pressure
transducers), and an all-metal mass-flow regulator to control the gas flow
into the OP cell. Downstream of the OP cell, the polarized gas is in contact
with non-magnetic materials only (glass, Teflon, low-degassing plastic tubes,
etc.). Leak-tight connections with respect to room air are required to avoid
$\mathrm{O}_{2}$-induced relaxation \cite{SaamO2}. Dead volumes are kept
as small as possible, with the constraint
that excess relaxation can result from drastic enhancement of
surface-to-volume ratios.

\begin{figure}[h]
\includegraphics[clip,width=3.15 in]{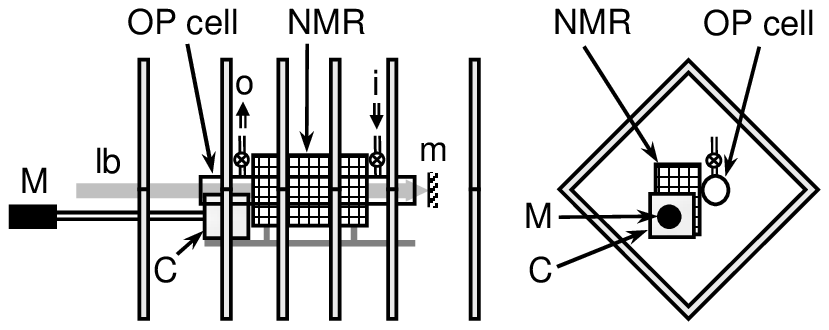} 
\caption{Sketch of the $^{3}\mathrm{He}$ hyperpolarization system (front and side
views), showing the arrangement of OP cell, compressor (C), driving motor (M),
and of the six square coils (average side length 40~$\mathrm{cm}$, section
1.6$\times$1.6~$\mathrm{cm}^{2}$) that provide the static field
(0.8516~$\mathrm{mT/A}$\textrm{\ }at the centre). Field map is optimized both
for OP and NMR by coil positions ($\pm$5.9, $\pm$19.1, and $\pm$%
37.8~$\mathrm{cm}$) and number of turns (85, 100, and 224, respectively).
2$\times$30~additional turns are wired in opposite directions onto the outer
square coils, yielding a 1.3~$\mathrm{mT/cm/A}$ shimming gradient over the
sample. A mirror (m) back-reflects the laser beam (lb) into the OP cell, open
at both ends for gas inlet (i) and outlet (o).}
\label{fig3}
\end{figure}

A weak static magnetic field (2~$\mathrm{mT}$) is generated by a set of six 
square coils of horizontal axis, designed to provide an optimal field
homogeneity both for OP and NMR in a compact setup (Fig.~\ref{fig3}).
Relative
inhomogeneities of the applied field do not exceed a few parts per thousand
over the entire cell volume for OP, and a few ppm over a 7-$\mathrm{cm}$
diameter spherical volume for NMR. The stray field inhomogeneity induced by
the laboratory environment is weak, and local compensation of the dominant
residual gradient is achieved using two shimming coils.

Absolute measurements of the nuclear polarization in the OP cell are performed
by optical polarization analysis of the fluorescence light emitted by the
plasma. With 2-$\mathrm{W}$ laser power, nuclear polarizations up to 60\% can
be obtained with no gas flow, and within the 30\%$-$40\% range around
1~$\mathrm{sccm}$ (standard cubic centimeter per minute)(0.745~$\mathrm{\mu mol/s}$) flow rates, mostly due to
pressure increase in the OP\ cell. The polarization losses in the compressor
and connecting tubes, characterized in independent measurements \cite{JC02},
typically amount to 10\%, at best.

\subsection{Probe gas management}

The filling tube attached to the aerogel sample is connected to a Luer
polycarbonate stopcock valve located just outside the NMR shielding box.
There, pressure (relative to the atmospheric pressure) is measured
with a 1\% accuracy in the 1~$\mathrm{mbar}-$1~$\mathrm{bar}$ range using a
piezoresistive sensor mounted in a nonmagnetic (plastic) case (Model
24PCCFA2G, Honeywell, Morriston NJ, USA). For low-pressure diffusion
measurements, pure $^{3}\mathrm{He}$ is used. The gas flows directly from the
OP cell and is accumulated inside the aerogel by the compressor. For high-pressure 
diffusion measurements, accumulation of large quantities of polarized
gas is of limited interest considering the bottleneck set to magnetization
buildup by wall relaxation (see Section~\ref{section3.2}). Therefore,
$\mathrm{N}_{2}$ is used as a neutral, nonrelaxing buffer gas to slow down the
diffusion of the $^{3}\mathrm{He}$ atoms in the void spaces of the aerogels.
The $^{3}\mathrm{He}$ partial pressures vary from 10 to 200~$\mathrm{mbars}%
$\ in the measurements, but most frequently no more than 30~$\mathrm{mbars}$ of
polarized gas are introduced in the sample containers (i.e., only a few
$\mu$%
moles). Total pressures range from 10~$\mathrm{mbars}$ to 1~$\mathrm{bar}$.

Before the NMR acquisition, the sample is flushed with pure $\mathrm{N}_{2}$
(99.995\% grade), then evacuated. Filling with polarized $^{3}\mathrm{He}$
using the peristaltic compressor takes a few tens of seconds. Pure
$\mathrm{N}_{2}$ is optionally added to reach the targeted pressure value,
then the measurement is performed. Right after signal acquisition, the gas is
removed and the aerogel is kept under vacuum until the next measurement. Each
series of measurements can last from a few hours to several days. When it is
completed, the sample is open to room air and stored at ambient pressure.

\section{NMR\ measurements}

\subsection{FID signals and line shapes}

The performed measurements do not require an absolute calibration of the NMR
detection system. However, we have used a small transmitting rf\ coil located
at the center of the setup to probe the global pick-up sensitivity, and also
to check that the $^{3}\mathrm{He}$ signals correspond to nuclear polarization
rates in agreement with those produced by the optical pumping system. The
measured noise level is about 0.7-$\mathrm{\mu V}$ rms when the LIA time
constant is set to minimum (160~$\mathrm{\mu s}$). Nuclear polarizations vary
from a few percents to 20\% at best, due to relaxation in the aerogel, and the
best spin-echo signal-to-noise ratios (SNRs) achieved are on the order of
30$-$50. Higher polarizations might be obtained if polarized gas is
accumulated in a small glass cell, then rapidly expanded into the sample
container. However, the procedure used here seems more suited to systematic
and repeated measurements in the fragile silica substrates, and it still
yields satisfactory accuracy.

In our experiments, FID signal lifetime is limited by residual field
inhomogeneities due to the magnetic environment of the experiment after
shimming (Section~\ref{section2.3}). The motional narrowing regime can then be
fully reached for low pressure gas, and exponential decay is observed with a
time constant $T_{2}^{\ast}$ exceeding 2.5~$\mathrm{s}$ at 60-$\mathrm{mbar}$
pressure (whereas $T_{2}^{\ast}<$0.5~$\mathrm{s}$ without shimming). At
1-$\mathrm{bar}$ pressure (no motional averaging) signal lifetime remains
shorter, typically on the order of 0.2~$\mathrm{s}$.

For the diffusion measurements, a strong gradient is applied so that at all
pressures the FID lifetimes become much shorter and the line shapes, obtained
by fast Fourier transform (FFT) of the NMR\ signals, are those imposed by the
one-dimensional (1D) spatial distribution of resonance frequencies (see Section~\ref{section3.3}%
). Otherwise, line distortions occur due to diffusive relaxation or spectral
edge enhancement effects \cite{Putz92,Callaghan93} and a variety of spectral
profiles can be observed, depending on gradient amplitude (i.e., frequency
span over the sample size) and gas pressure (i.e., diffusion time during
observation) conditions. Particularly important in highly diffusive systems,
these line distortion effects have recently been systematically characterized
with good SNR using hyperpolarized noble gases \cite{Saam96,Song98}. Care has
thus been taken to select gradient strengths and echo times so as to avoid
systematic biases in the diffusion measurements
\cite{Wayne66,Putz92,deSwiet94,Hurlimann95,Hayden04}.

\subsection{Longitudinal relaxation rates\label{section3.2}}

The contribution of dipole-dipole relaxation to the longitudinal relaxation
time $T_{1}$ is totally negligible for pure $^{3}\mathrm{He}$ (relaxation time
inversely proportional to the $^{3}\mathrm{He}$ pressure, exceeding
800~$\mathrm{h}$ at 1~$\mathrm{bar}$ \cite{T1He3}). Only severe contamination
by air may play a significant role in our experiments, through strong
dipole-dipole relaxation by the paramagnetic $\mathrm{O}_{2}$ molecules
(12.5$~\mathrm{s}$ at 1~$\mathrm{bar}$ and 21\% oxygen concentration
\cite{SaamO2}). Bulk relaxation due to atomic diffusion in our residual field
gradients is also negligible. Wall relaxation is thus the dominant process,
and the magnetization lifetime depends on the nature of the materials in
contact with the gas, and on the area of exposed surfaces.

The longitudinal relaxation time $T_{1}$ is measured to be 370~$\mathrm{s}$
in the empty machined PMMA\ cell. This corresponds to a specific relaxation
time $\theta_{s}=T_{1}A/V$ ($A$: area, $V$:\ volume) equal to $0.19\times
10^{6}~\mathrm{s/m}$, slightly lower than typical values reported for Pyrex
glass ($\theta_{s}=0.5-3.8\times10^{6}~\mathrm{s/m}$
\cite{Fitzsimmons69,Heil95,Jacob03}) and quartz ($\theta_{s}=0.3-2.8\times
10^{6}~\mathrm{s/m}$ \cite{Fitzsimmons69,PJNunpub}).

Surprisingly long $T_{1}$ are measured in the aerogels, especially in the
commercial ones, despite their huge nominal specific areas (a few tens of
$\mathrm{m}^{2}$ per $\mathrm{cm}^{3}$) and uncontrolled cleanliness. $T_{1}$
reaches 130$~\mathrm{s}$ and 200$~\mathrm{s}$ in samples A and J$_{3}$,
respectively. The contribution of the container walls can be estimated from
the reference PMMA cell $T_{1}$. This yields particularly weak specific
relaxation times on the silica aggregates ($\theta_{s}=6.6$ and $25\times
10^{9}~\mathrm{s/m}$, respectively), two orders of magnitude longer than the
record values obtained with special aluminosilicate glasses (low-boron content
Corning 1720 glass: $2.0\times10^{7}~\mathrm{s/m}$ \cite{Gentile01}; Schott
iron-free Supremax glass: 1.6$-$4.2$\times10^{7}~\mathrm{s/m}$ %
\cite{Heil99,WHprivate}). $T_{1}\ $does not exceed $26~\mathrm{s}$ in the
custom-made sample M, which is less than initially measured at
2.3~$\mathrm{mT}$ (58~$\mathrm{s}$) but more than subsequently obtained at
0.1~$\mathrm{T}$\ (19~$\mathrm{s}$) in the same sample \cite{GT00}.
Ferromagnetic impurities in the material
\cite{Mazitov93,Fitzsimmons69,Jacob03} may be responsible for these hysteretic
$T_{1}$ changes, also observed recently in bare glass cells exposed to high
magnetic fields \cite{WHprivate} (a phenomenon reported for rubidium-coated
glass cells only, so far \cite{Jacob01,Jacob04}). Using an estimate for the
Pyrex $T_{1}$, and an approximate specific area (22.9$~\mathrm{m}%
^{2}\mathrm{/cm}^{3}$ for a $98.2\%$\ similar custom-made aerogel
\cite{Porto96}), the 58-$\mathrm{s}$ $T_{1}$ initially measured in sample M
would still yield $\theta_{s}=1.5\times10^{9}~\mathrm{s/m}$ for the silica
aggregates (shorter than, but comparable to, that of the commercial aerogels).
In all samples, magnetization decays are indeed slow enough to allow the
one-shot CPMG diffusion measurements described hereafter.

\subsection{Spin-echo attenuation\label{section3.3}}

A $90_{x}^{\circ}$ rf pulse ($0.5$-$\mathrm{ms}$ duration) is followed by a
train of $180_{y}^{\circ}$ rf pulses ($1$-$\mathrm{ms}$ duration) to refocus
the transverse magnetization at a regular interval $T_{\mathrm{CP}}$. CPMG
measurements are made for different values of $T_{\mathrm{CP}}$ (6$-$%
31~$\mathrm{ms}$) and of applied gradient $G$ (0$-$2~$\mathrm{mT/m}$).
Spatially nonselective tipping is obtained with a good angle accuracy by
switching off (on) the applied gradient $\mathrm{300~\mu s}$ before (after)
the beginning (the end) of the rf\ pulses.

For the CPMG sequence, the transverse magnetization has a decay time
$T_{2}^{\prime}$ that can be written as
\begin{equation}
\frac{1}{T_{2}^{\prime}}=\frac{1}{T_{2}}+\frac{1}{T_{2,\mathrm{diff}}}
\label{eq:t2general}%
\end{equation}
The first term on the right-hand side is the intrinsic relaxation rate of the
system, that for the gas simply corresponds to the longitudinal decay rate
($T_{2}=T_{1}$). This has been checked in the absence of applied gradient,
using spin-echo trains with very short $T_{\mathrm{CP}}$. The second term
represents additional relaxation resulting from diffusion in a nonuniform
magnetic field, due to an applied field gradient and/or the residual magnetic
inhomogeneities after shimming.

For a uniform applied gradient $G$, $T_{2,\mathrm{diff}}$ can be expressed in
the case of free gas diffusion as
\begin{equation}
T_{2,\mathrm{diff}}=\frac{12}{D(\gamma kGT_{\mathrm{CP}})^{2}}
\label{eq:tdiff}%
\end{equation}
where $D$ is the coefficient of diffusion and the factor $k$ corrects for the
time variation of the applied gradient, which is switched off during the rf
pulses. For a rectangular gradient pulse, this correction factor is:
\begin{equation}
k=\sqrt{\delta^{2}(3T_{\mathrm{CP}}-2\delta)/T_{\mathrm{CP}}^{3}}
\label{eq:Gcorr}%
\end{equation}
where $\delta\leq T_{\mathrm{CP}}$ is the duration of the gradient
\cite{Stejskal}.

\begin{figure}[h]
\includegraphics[clip,width=3.15 in]{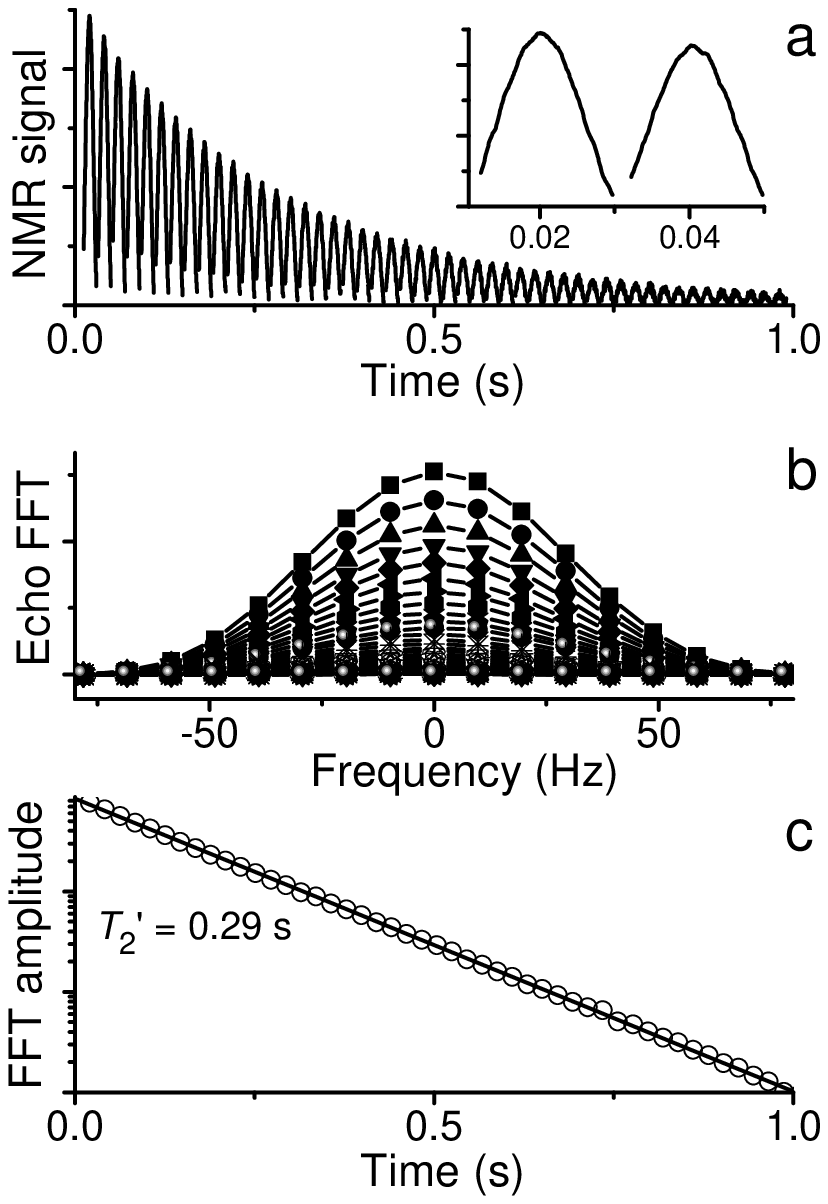} 
\caption{a: Example of spin-echo train obtained in aerogel A (with 33.4~$\mathrm{mbars}$
of $^{3}\mathrm{He}$ plus 616.5~$\mathrm{mbars}$ of $\mathrm{N}_{2}$,
$G=0.314$~$\mathrm{mT/m}$, and $T_{\mathrm{CP}}=21~\mathrm{ms}$). Saturation of
the detection system by the strong rf pulses occurs between echoes (data
points removed). Inset: Blow-up plot of first and second echoes. b:
Fast Fourier transforms (FFTs) of the consecutive echo data sets, identified by
individual symbol shapes. c: Exponential
fit of FFT decay, yielding $T_{2}^{\prime}=$ 0.29~$\mathrm{s}$.}
\label{fig4}
\end{figure}

All diffusion-weighted spin-echo trains are observed to decay exponentially.
An example of spin-echo train and FFT is presented in Fig. \ref{fig4}.
$T_{2}^{\prime}$ values are obtained from monoexponential fits to the square of the
echo amplitudes, or the square of their Fourier components. The use of power
data, rather than magnitudes, yields a more reliable result as they are
less sensitive to biases introduced by noise base lines \cite{weerd}.

\begin{figure}[h]
\includegraphics[clip,width=3.15 in]{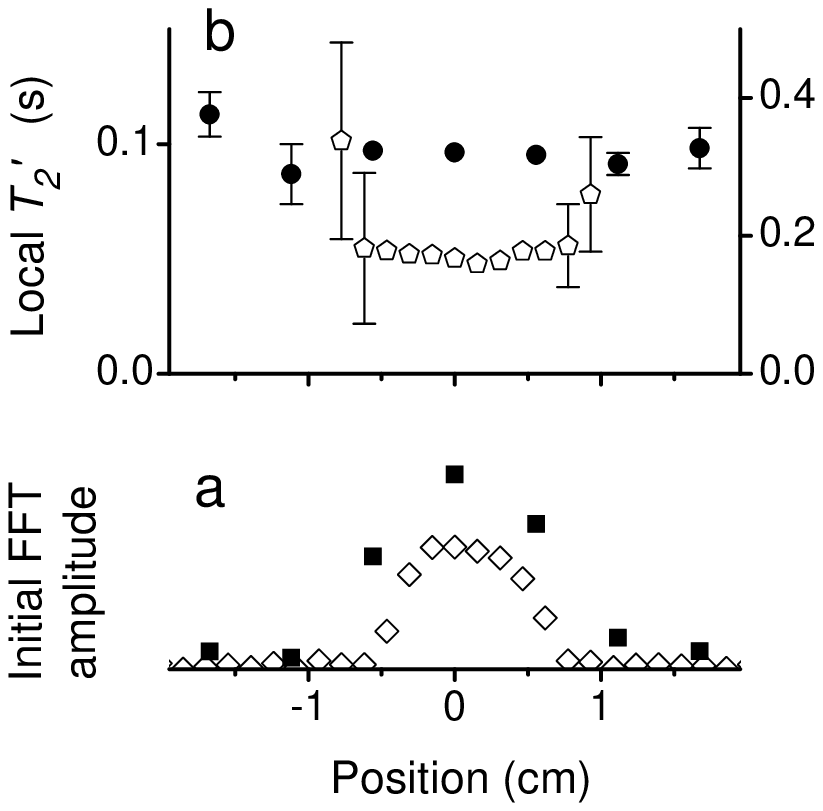} 
\caption{Pixel-per-pixel analysis of frequency spectra, through monoexponential fits
of the FFT decays, for spin-echo trains obtained in aerogels A (open symbols)
and J$_{2}$ (solid symbols) with $G$=1.56 and 0.43 ~$\mathrm{mT/m}$,
respectively, and $T_{\mathrm{CP}}$=11~$\mathrm{ms}$. a: Local initial
amplitudes (along gradient axis). Statistical error bars, not plotted, are
less than\ 2\% of the maximum amplitude. b: Local decay times (left axis: A;
right axis: J$_{2}$). Statistical
error bars are smaller than symbol sizes, except where plotted (i.e., on
sample edges).}
\label{fig5}
\end{figure}

The diffusion measurements are typically performed with short repetition times
($T_{\mathrm{CP}}$= 6 or 11~$\mathrm{ms}$) and finite gradients ($G$%
=0.4$-$1.2~$\mathrm{mT/m}$), and $T_{2}^{\prime}$ ranges from 25 
to 1500~$\mathrm{ms}$ in aerogel (25 to 86~$\mathrm{ms}$ in the reference
cell). The contribution of intrinsic $T_{2}$ to transverse relaxation is thus
negligible at all pressures, and $T_{2}^{\prime}=T_{2,\mathrm{diff}}$. The
frequency spectra of the recorded echoes provide 1D images of the gas
magnetization in the samples (Fig.~\ref{fig5}a). 
The
shape of sample A is fairly well reproduced considering the selected spatial
NMR resolution 2$\pi/(\gamma GT_{\mathrm{obs}})$=2.5~$\mathrm{mm}$ ($\gamma
$=2$\pi\times$32.43~$\mathrm{MHz/T}$ is the gyromagnetic ratio of
$^{3}\mathrm{He}$ nuclei, and $T_{\mathrm{obs}}$ the observation time that is
very close to the gradient duration $\delta$). Resolution is lower for the
displayed J$_{2}$ profile (7.8~$\mathrm{mm}$), but typical of most
measurements presented in Section~\ref{section4}. The frequency spectra of the
echoes also provide 1D maps of the decay time (Fig.~\ref{fig5}b). Spatial
variations ($\pm$10\%) are small enough to accurately yield a single
exponential $T_{2}^{\prime}$ for the average magnetization.

\begin{figure}[h]
\includegraphics[clip,width=3.15 in]{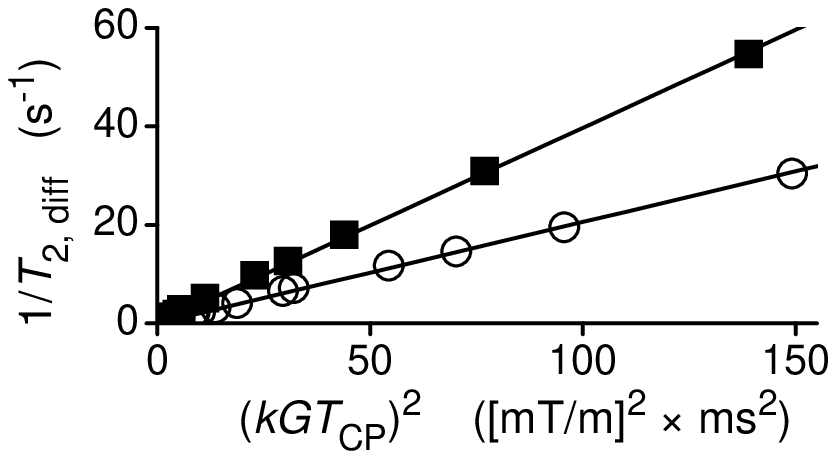} 
\caption{Compilation of spin-echo decay rates measured in the reference cell
(solid squares) and in aerogel M (open circles) at fixed gas pressures.
Diffusion-weighting parameter $(kGT_{\mathrm{CP}})^{2}$ is varied through
changes of gradient strengths ($G$: $0.07-2.2$
$\mathrm{mT/m}$) and echo times ($T_{\mathrm{CP}}$: $4.4-31~\mathrm{ms}$).
The expected linear
dependence [Eq.~(\ref{eq:tdiff})], observed over more than two decades, yields
$D_{\mathrm{gas}}=1.148\pm0.004$ $\mathrm{cm}%
^{2}\mathrm{/s}$ in the reference cell, and
$D_{\mathrm{eff}}=0.595\pm0.008$ $\mathrm{cm}^{2}\mathrm{/s}$ in aerogel M,
for the two $^{3}\mathrm{He}$-$\mathrm{N}_{2}$ mixtures used
($P_{3}^{\mathrm{eq}}=1.41$~$\mathrm{bar}$ and
$1.70$~$\mathrm{bar}$, respectively - see Section~\ref{section4.1}).}
\label{fig6}
\end{figure}

From each measurement, a diffusion coefficient $D$ is extracted using
Eq.~(\ref{eq:tdiff}). Experimentally, the $1/T_{2}^{\prime}$ data can be checked
to exhibit the expected linear variation with $(kGT_{\mathrm{CP}})^{2}$ at all
pressures above 100~$\mathrm{mbars}$. Fig.~\ref{fig6} displays, for instance,
two series of measurements performed in the reference cell and in aerogel M at
fixed gas pressures. 
Various
combinations of parameters $G$ and $T_{\mathrm{CP}}$ are used here, and in both cases
all the data collapse on a straight line. In the empty reference cell, $D$
coincides with the free-diffusion coefficient $D_{\mathrm{gas}}$ of the
$^{3}\mathrm{He}$ atoms in the probe gas. In aerogel, $D$ corresponds to an
effective diffusion coefficient $D_{\mathrm{eff}}$ that describes the
restricted diffusion of the atoms within the porous medium. Below
100~$\mathrm{mbars}$, moderate departures from the $(kGT_{\mathrm{CP}})^{2}$
dependence can be observed in aerogel at high $GT_{\mathrm{CP}}$,
corresponding to reduced decay rates $1/T_{2}^{\prime}$ (hence, overestimated
values of $D$), typically above $kGT_{\mathrm{CP}}=4\times10^{-6}~\mathrm{sT/m}$ as
reported in Ref. \cite{GG01}.

\section{Results\label{section4}}

\subsection{Free gas diffusion\label{section4.1}}

The free-diffusion coefficient $D_{\mathrm{gas}}$ for a $^{3}\mathrm{He}$ gas
with partial pressure $P_{3}$ in a mixture containing $\mathrm{N}_{2}$ with
partial pressure $P_{\mathrm{N2}}$ can be written as:
\begin{equation}
\frac{1}{D_{\mathrm{gas}}}=\frac{P_{3}}{\mathcal{D}_{3}(T)}+\frac
{P_{\mathrm{N2}}}{\mathcal{D}_{3\mathrm{N2}}(T)},\label{eq D3He}%
\end{equation}
where the reduced diffusion coefficients can be expressed as $\mathcal{D}%
_{3}=1.967\times\left(  T/300\right)  ^{1.71}$ and $\mathcal{D}_{3\mathrm{N2}%
}=0.811\times\left(  T/300\right)  ^{1.65}$ in units of $\mathrm{atm}%
\times\mathrm{cm}^{2}\mathrm{/s}$ (1~$\mathrm{atm}=1.013\times10^{5}%
$~$\mathrm{Pa}$), with the temperature $T$ in $\mathrm{Kelvin}$
\cite{Bidinosti03,note1}. Therefore, the scattering on $\mathrm{N}_{2}$
molecules leads to an effect of the gas composition on atomic diffusion that
can be accounted for using the \textit{equivalent pressure} of pure
$^{3}\mathrm{He}$ gas:
\begin{equation}
P_{3}^{\mathrm{eq}}=P_{3}+P_{\mathrm{N2}}\frac{\mathcal{D}_{3}(T)}{\mathcal{D}%
_{3\mathrm{N2}}(T)},\label{eq Peq}%
\end{equation}
with $\mathcal{D}_{3}/\mathcal{D}_{3\mathrm{N2}}=3.462$ at 293~$\mathrm{K}$.
This allows to merge the results obtained in $^{3}\mathrm{He}-\mathrm{N}_{2}$
mixtures and in pure $^{3}\mathrm{He}$ gas. For free diffusion,
$1/D_{\mathrm{gas}}$ and hence $T_{2}^{\prime}$ are expected to linearly
depend on $P_{3}^{\mathrm{eq}}$.

\begin{figure}[h]
\includegraphics[clip,width=3.15 in]{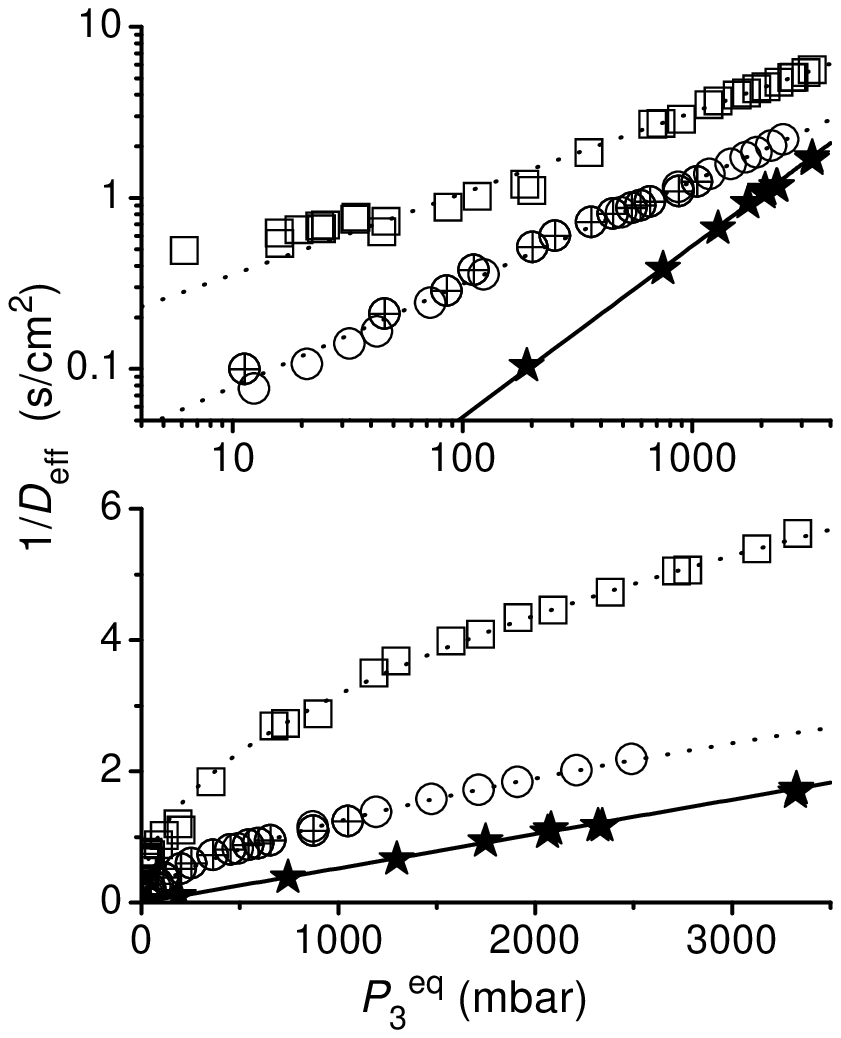} 
\caption{Comparison of diffusion measurements performed in free gas, and in aerogels A
and M. The same three sets of reduced data 1/$D_{\mathrm{eff}}$, plotted
against pressure $P_{3}^{\mathrm{eq}}$, are displayed
 in linear and logarithmic scales. Stars: reference cell
data, free diffusion. Circles (respectively, squares): aerogel M (resp. A) data.
Crossed circles: previous measurements\ in aerogel M \cite{GT00b,GG01}.
Solid line: linear fit of the free-diffusion
data (slope: 5.06$\pm$0.04 $\mathrm{mbar}^{\mathrm{-1}}\times\mathrm{s/m}%
^{2}\mathrm{)}$, in good agreement with the expected $P_{3}^{\mathrm{eq}}%
/\mathcal{D}_{3}$ dependence. Dotted lines: phenomenological
power-law fits with exponents 0.47 (aerogel A) and 0.60 (aerogel M).}
\label{fig7}
\end{figure}

CPMG\ measurements performed in the reference cell are presented in
Fig.~\ref{fig7}. These data are obtained at various $^{3}\mathrm{He}$
pressures, with several combinations of parameters $G$ and $T_{\mathrm{CP}}$. They
actually all collapse on a straight line when plotted as a function of the
equivalent $^{3}\mathrm{He}$ pressure $P_{3}^{\mathrm{eq}}$. Moreover, the measured
slope $5.06\pm0.04~\mathrm{mbar}^{-1}\times\mathrm{s/m}^{2}$ (statistical
error bar) is in fair agreement with the expected value ($1/\mathcal{D}%
_{3}=5.23~\mathrm{mbar}^{-1}\times\mathrm{s/m}^{2}$) given the 4\% uncertainty
due to long-term fluctuations of atmospheric pressure and
temperature. This provides a consistency check for our apparatus, gradient
assignment, and measurement technique. 

\subsection{Gas diffusion in aerogel}

\begin{figure}[h]
\includegraphics[clip,width=3.15 in]{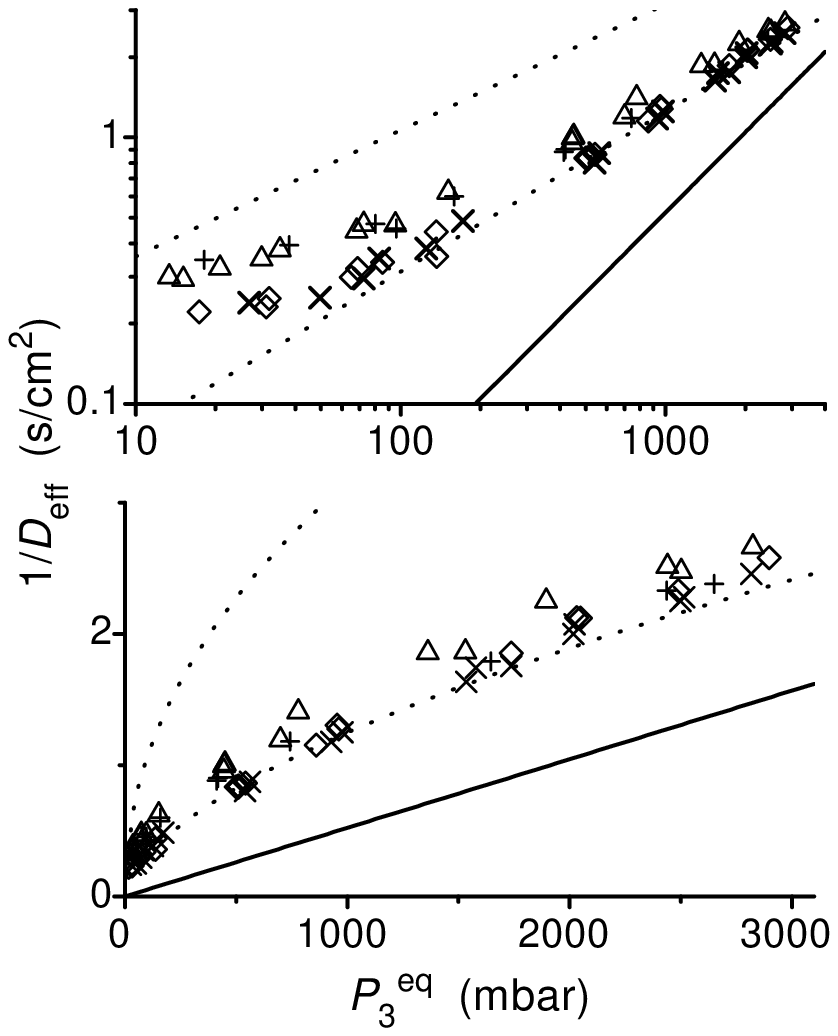} 
\caption{Reduced diffusion data 1/$D_{\mathrm{eff}}$ as a function of $P_{3}^{\mathrm{eq}}$ for
aerogel samples J$_{1}$ (plus signs), J$_{2}$ (open triangles),
J$_{3}$ (open diamonds), and J$_{4}$ (crosses). Comparison is made
with the linear fit of free-diffusion 
data (straight line), and power-law fits of aerogels A and M data
from Fig.~\ref{fig7} (dotted lines). The J$_{1}-$J$_{4}$ data can be
consistently fitted by a power-law functions 
with exponents 0.53 or 0.64 (see Table~\ref{table1}).}
\label{fig8}
\end{figure}

Results obtained in aerogel samples A and M are also displayed in
Fig.~\ref{fig7}, so that they can be directly compared with the free-diffusion
results. For clarity, the results obtained in samples J$_{1}-$J$_{4}$ are
presented separately in Fig.~\ref{fig8}. 

In Fig.~\ref{fig7}, data from earlier measurements in sample M are also
included for comparison. These
measurements \cite{GT00b,GG01} have been performed in a 0.1~$\mathrm{T}$ whole-body imager,
using the broadband rf channel and the gradient system of that clinical unit.
Homemade antennas were used for induction and detection at 3.24~$\mathrm{MHz}%
$, as well as the in-house $^{3}\mathrm{He}$-NMR software for sequence
management and data acquisition developed for \textit{in vivo} lung studies
\cite{cras97,Durand02}. Pure $^{3}\mathrm{He}$ gas was used for these CPMG\ measurements
($G=0.21~\mathrm{mT/m}$, $T_{\mathrm{CP}}=10^{-2}~\mathrm{s}$). Despite all
these differences in apparatus and experimental conditions, the two sets of
data are perfectly consistent. The present work significantly extends the
probed range of gas pressures. It confirms both the global decrease in
diffusion coefficient (as compared to free diffusion) and the strong nonlinear 
dependence on gas pressure. These results therefore appear to be quite
robust, and reliably associated to intrinsic bulk characteristics of the
aerogel (on which time or rough pressure variations seem to have no effect).

In Fig.~\ref{fig7}, the results obtained in the other aerogel (A) are
qualitatively similar but show a strikingly stronger departure from the free-diffusion 
behavior, revealing a much more restricted $^{3}\mathrm{He}$
diffusion. This may be attributed to the difference in aerogel porosities, the
silica volume density being twice larger in A than in M. The quantitative
difference in pressure dependences is emphasized by the logarithmic plot, that
shows that the two sets of data are best fitted with distinct power laws.

Type J aerogel samples have exactly the same nominal porosity as A, and yet
the data for J$_{1}-$J$_{4}$ are actually much closer to those of M (see
Fig.~\ref{fig8}). Finally, data obtained in the four samples of type J are
quite satisfactorily comparable, revealing only minor sample-to-sample
variations (again, there is no sign of individual alteration by inflicted
mechanical stress). These data are of lower quality, the SNR being lesser due
to the smaller sizes of the J$_{1}-$J$_{4}$ samples.

$1/D_{\mathrm{eff}}$ exhibits a strong nonlinear pressure dependence in all
aerogels. The logarithmic graphs in Figs.~\ref{fig7} and \ref{fig8} show that
the data can be described by simple power-law functions, with exponents lying
in the 0.48$-$0.65 range. For aerogels A and J$_{1}-$J$_{4}$, the data depart from
that common behavior (longer $T_{2}^{\prime}$ are measured) below
100~$\mathrm{mbars}$, and this value is chosen as a lower bound for the fitting
range. At the lowest pressures, limits of the validity domain of
Eq.~(\ref{eq:tdiff}) (see Section~\ref{section3.3}) are indeed reached, due to
the experimental conditions for measurements performed in these samples
($10^{6}kGT_{\mathrm{CP}}=4.5$ and $6.1~\mathrm{Ts/m},$ respectively). With
the choice made for sample M ($10^{6}kGT_{\mathrm{CP}}=3.0~\mathrm{Ts/m),}$ no
such departure occurs down to 10~$\mathrm{mbars}$ and all data are included in
the power-law fit. Fit parameters are listed in Table~\ref{table1} to sum up
sample-to-sample variations. 
\begin{table*}[ht]
\centering%
\begin{tabular}
[c]{|c|c|c|c|c|c|}\hline
Sample & Manufacturer & Porosity & Specific area & Exponent & Prefactor
[value for $P_{3}^{\mathrm{eq}}=$1 bar]\\
&  & [\%] & [m$^{2}$/g] & $\beta$ (stat. err.) & $\alpha$ (stat. err.)
[s/cm$^{2}$]\\\hline
M & custom made & 98.5 & - & 0.60 (0.01) & 1.25 (0.01)\\\hline
A & Airglass & 97 & 465 & 0.47 (0.01) & 3.16 (0.03)\\\hline
J$_{1}$ & Matsushita & 97 & 510 & 0.53 (0.02) & 1.42 (0.02)\\\hline
J$_{2}$ & Matsushita & 97 & 510 & 0.53 (0.02) & 1.54 (0.02)\\\hline
J$_{3}$ & Matsushita & 97 & 510 & 0.64 (0.01) & 1.31 (0.01)\\\hline
J$_{4}$ & Matsushita & 97 & 510 & 0.64 (0.02) & 1.27 (0.02)\\\hline
\end{tabular}
\caption{Characteristics and fit parameters for the six aerogel samples.
Exponents ($\beta$) and prefactors ($\alpha$) are obtained by
phenomenological fits of the NMR diffusion data to 1/$D_{\mathrm{eff}}$
$=\alpha \times (P_{3}^{\mathrm{eq}})^{\beta}$, where 1/$D_{\mathrm{eff}}$ is in
$\mathrm{s/cm}^{\mathrm{2}}$ and $P_{3}^{\mathrm{eq}}$ in $\mathrm{bars}$ (see
text).}%
\label{table1}%
\end{table*}

\section{Discussion}

The use of hyperpolarized $^{3}\mathrm{He}$ is demonstrated to allow diffusion
measurements with good accuracy, over a wide range of gas densities. The
compact polarizer is perfectly suited to online preparation of the small gas
quantities required to test the aerogel samples, and systematic measurements
are easy to carry out and repeat. The high porosity samples under study are
selected among silica aerogels that are most frequently used as host porous
matrices for phase transition studies in low-temperature physics. None of
these particular samples has been submitted to cool down-warm up, or
compression-decompression, cycles comparable to those routinely performed in
the low-temperature experiments. Exposure to high pressure or thermal stress
has recently been demonstrated to induce irreversible changes in aerogels,
that may macroscopically affect their structure \cite{Eska03}. It might thus
be interesting to perform nondestructive NMR diffusion measurements in
aerogel both before and after use at low temperature.

The consistent set of robust NMR data collected in the differently produced
virgin samples reveals common bulk aerogel features, that are analyzed in
terms of specific properties of the microscopic structure in this section. The
unusually high porosities of the samples allow a particularly simple
description of the gas kinetics inside the aerogels. Restricted diffusion is
more often studied in denser media, where atoms or molecules are confined
inside an ensemble of closed or interconnected pores that strongly constrains,
by its geometry or tortuosity, the thermal motion of the probe species. In
this work, the solid matrix occupies a very small fraction ($1.5\%$ to $3~\%$)
of the volume of the aerogel, and the hyperpolarized $^{3}\mathrm{He}$ atoms
diffuse inside the widely accessible, huge space left open by the sparsely
distributed silica aggregates.

Decay times of spin-echo amplitudes are measured to be two to five times longer in
aerogel than in free gas. The decrease of the atomic diffusion coefficient is
quite significant at pressures exceeding 1~$\mathrm{bar}$, indicating that
atom-atom and atom-silica collisions have comparable frequencies at these high
gas densities. Surprisingly, a strong pressure dependence of $D_{\mathrm{eff}%
}$ is still observed at low pressures. As the probability of atom-atom
scattering vanishes in a very dilute gas, the relative effect of pressure
variations on the $^{3}\mathrm{He}$ diffusion coefficient would be expected to
become increasingly small at low pressures (and some characteristic length
scale, if any, of the silica network should determine the finite value
approached by the atom mfp at zero pressure - see Section~\ref{section5.1}%
). It is yet measured to remain large at low pressures, indicating that the
free motion of the atoms is indeed allowed over large distances in aerogel (at
10~$\mathrm{mbars}$, for instance, the mfp for free diffusion in
$^{3}\mathrm{He}$ is on the order of 40~$\mathrm{\mu m}$ \cite{lpm!}). This is
systematically observed, in the custom-made aerogel sample and in the set of
commercial ones that are obtained from two distinct manufacturers. It brings
further support to the suggested nonuniform distribution of density of silica
strands inside the aerogels, mentioned in the Introduction, that had been put
forward to explain some preliminary results \cite{GT00b,GG01}.

The NMR gas diffusion data actually have two distinct features that compel
description of the aerogel by more than a single characteristic length: the
dominant power-law increase with pressure, and the departure from it that can
be noticed at very low pressures. This departure has already been discussed in
Ref. \cite{GG01}, and shown to be enhanced at large $kGT_{\mathrm{CP}}.$ As
pressure decreases, the atomic diffusion gets more and more sensitive to
scattering by the silica strands. This enhances the effects of\ spatial
distribution inhomogeneities and the multiexponential nature of the spin-echo
decay, as in other porous media \cite{Callaghan}. The longest lived mode is
favored for all length scales below the gradient-induced resolution, and
monoexponential fits systematically lead to notably underestimated relaxation
rates (i.e., increased 1/$D_{\mathrm{eff}}$ values). The following discussion
focuses on the power-law dependence on pressure observed in all samples and on
its analysis. Results are also compared to those obtained with other
investigation techniques.

\subsection{Crude model: uniform medium with a single characteristic
mfp\label{section5.1}}

The $^{3}\mathrm{He}$ diffusion process leading to spin-echo attenuation in
the applied magnetic-field gradient can be viewed as a random walk in
which free-flying atoms take successive individual steps that are terminated
either by a collision with another\ gas atom or molecule ($^{3}\mathrm{He}$ or
$\mathrm{N}_{2}$) or with silica.\ The frequencies of the two types of
collisions being additive, the mfp $\lambda_{\mathrm{eff}}$ of this random
walk can be related to the mfp $\lambda_{\mathrm{gas}}$ for gas collisions
and $\lambda_{\mathrm{aero}}$ for collisions with the silica strands in a
straightforward manner:
\begin{equation}
\frac{1}{\lambda_{\mathrm{eff}}}=\frac{1}{\lambda_{\mathrm{gas}}}+\frac
{1}{\lambda_{\mathrm{aero}}}\text{.} \label{eq1}%
\end{equation}
The spatial distribution of silica is here assumed to be homogeneous, and the
global effect of the aerogel structure on atomic motion is thus described by
the single parameter $\lambda_{\mathrm{aero}}$. The effective diffusion
coefficient $D_{\mathrm{eff}}$ can then be written as:
\begin{equation}
\frac{1}{D_{\mathrm{eff}}}=\frac{1}{D_{\mathrm{gas}}}+\frac{1}%
{D_{\mathrm{aero}}}\text{,} \label{eq2}%
\end{equation}
where the first contribution (gas-gas scattering) involves the free-diffusion
coefficient of $^{3}\mathrm{He}$ introduced in Section~\ref{section4.1}, and
the second one ($^{3}\mathrm{He}-$silica scattering) introduces the constant
diffusion coefficient $D_{\mathrm{aero}}=v\lambda_{\mathrm{aero}}/3$, where
$v$ is the average thermal velocity ($v=\sqrt{8k_{B}T/\pi m}$; $k_{B}$:
Boltzmann constant, $m$: atomic mass, $T$: temperature). The quantity
$1/D_{\mathrm{eff}}$ should therefore increase linearly with the gas pressure
(or the equivalent pressure $P_{3}^{\mathrm{eq}}$ for a $^{3}\mathrm{He-N}_{2}$
mixture), starting from a finite zero-pressure value that may vary from sample
to sample with the degree of confinement of the $^{3}\mathrm{He}$ atoms by the
silica network.

Over the wide range of the gas pressures investigated, the expected linear
dependence on pressure is clearly not observed. This strongly rules out the
description by a uniform density of silica scattering centres in the aerogels
and an average mfp. Instead, a broad distribution of mean free paths at
scales smaller than the imaging resolution has to be assumed to account for
the experimental observations. This is quantitatively discussed in
the following (Section~\ref{section5.2}), where the main features of required mfp
variations are correlated, over the probed range (0.1$-$40~$\mathrm{\mu
}\mathrm{m}$), with the measured pressure dependences.

\subsection{Phenomenological model: continuous distribution of mfp's in
aerogel\label{section5.2}}

To describe diffusion in heterogeneously distributed silica, a continuous
distribution $f$ of mfp's is introduced. The probability that the
$^{3}\mathrm{He}$-silica mfp $\lambda_{\mathrm{aero}}$ lies
between $\lambda_{\mathrm{aero}}$ and $\lambda_{\mathrm{aero}}+d\lambda
_{\mathrm{aero}}$ in some volume fraction of the sample is $f(\lambda
_{\mathrm{aero}})d\lambda_{\mathrm{aero}}$. Assuming that the atoms rapidly
exchange between regions of space characterized by different mfp's, the
$^{3}\mathrm{He}$\ magnetization is uniform and decays with a weighted average
of local relaxation rates:
\begin{equation}
\frac{1}{T_{2,\mathrm{diff}}}=\int\frac{f(\lambda_{\mathrm{aero}}%
)d\lambda_{\mathrm{aero}}}{T_{2,\mathrm{diff}}(\lambda_{\mathrm{aero}}%
)}\text{.}\label{eq3}%
\end{equation}
Since the local relaxation times are given by:
\begin{equation}
T_{2,\mathrm{diff}}(\lambda_{\mathrm{aero}})=\frac{12}{(\gamma
kGT_{\mathrm{CP}})^{2}}(\frac{1}{D_{\mathrm{gas}}}+\frac{1}{D_{\mathrm{aero}%
}(\lambda_{\mathrm{aero}})}),\label{eq4}%
\end{equation}
the effective diffusion coefficient derived from the global decay time
$T_{2,\mathrm{diff}}$ using Eq.~(\ref{eq:tdiff}) reads:
\begin{eqnarray}
D_{\mathrm{eff}}=\int\frac{f(\lambda_{\mathrm{aero}})D_{\mathrm{aero}}%
(\lambda_{\mathrm{aero}})D_{\mathrm{gas}}d\lambda_{\mathrm{aero}}%
}{D_{\mathrm{gas}}+D_{\mathrm{aero}}(\lambda_{\mathrm{aero}})}\nonumber\\=\int
\frac{f(\lambda_{\mathrm{aero}})\mathcal{D}_{3}d\lambda_{\mathrm{aero}}%
}{P_{\mathrm{aero}}(\lambda_{\mathrm{aero}})+P_{3}^{\mathrm{eq}}}.\label{eq5}%
\end{eqnarray}
The physical quantity $P_{\mathrm{aero}}(\lambda_{\mathrm{aero}}%
)=3\mathcal{D}_{3}/v\lambda_{\mathrm{aero}}$, that appears in the right hand
side of Eq.~(\ref{eq5}), has the dimension of a pressure. It represents the
contribution of the $^{3}\mathrm{He}$-silica scattering process to hindering
of atomic motion, on the $\lambda_{\mathrm{aero}}$ length scale, by the solid
matrix. Eq.~(\ref{eq5}) shows that the integrand is a simple rational function
of the probe gas pressure. The variations of $D_{\mathrm{eff}}$ with
$P_{3}^{\mathrm{eq}}$ therefore put strong constraints on the mfp distribution $f$
introduced in this model.

Experimentally, $D_{\mathrm{eff}}$ is proportional to $(P_{3}^{\mathrm{eq}})^{-\beta}$
over a wide range of pressures. This is mathematically equivalent to set
$D_{\mathrm{eff}}(aP_{3}^{\mathrm{eq}})=a^{-\beta}D_{\mathrm{eff}}(P_{3}^{\mathrm{eq}})$, $a$
being is an arbitrary scaling constant. Through a straightforward change from
$\lambda_{\mathrm{aero}}$ to $a\lambda_{\mathrm{aero}}$ as dummy variable in
the integral [Eq.~(\ref{eq5})], this relation can be shown to hold for all
values of $P_{3}^{\mathrm{eq}}$ if, and only if, $f(\lambda_{\mathrm{aero}}/a)=a^{2-\beta
}f(\lambda_{\mathrm{aero}})$. This translates into $f(\lambda_{\mathrm{aero}%
})\varpropto(\lambda_{\mathrm{aero}})^{\beta-2}$. The mfp distribution,
itself, corresponds to a power law, and its exponent is directly related to
that measuring the $1/D_{\mathrm{eff}}$ data scaling rate versus pressure.
Although $P_{3}^{\mathrm{eq}}$ is varied over more than two decades, $D_{\mathrm{eff}}$
is actually measured over a finite pressure interval. Additional information
would thus be required to fully characterize $f(\lambda_{\mathrm{aero}}).$
Yet, the distribution $f(\lambda_{\mathrm{aero}})\varpropto(\lambda
_{\mathrm{aero}})^{\beta-2}$ provides a convenient tool for quantitative
analysis and discussion of the experimental results.

\begin{figure}[h]
\includegraphics[clip,width=3.15 in]{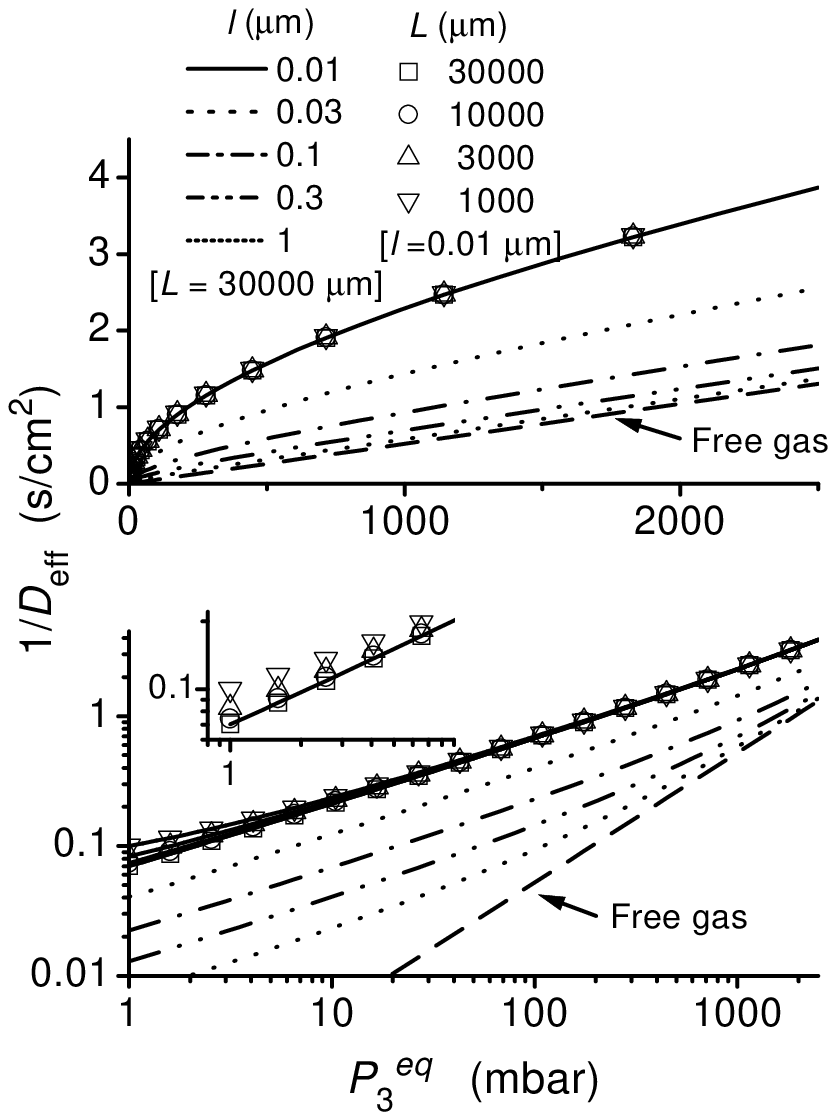} 
\caption{Pressure dependence of $1/D_{\mathrm{eff}}$ \ computed \ from Eq.~(\ref{eq5})
for broad power-law mfp distributions in aerogel $f(\lambda_{\mathrm{aero}%
})\varpropto(\lambda_{\mathrm{aero}})^{\beta-2}$ with $\beta=0.5$, as a
function of lower ($l$) or upper ($L$) bounds. The mfp distributions range
either from $l=$0.01~$\mathrm{\mu m}$ to $L$=1, 3, 10, and 30~$\mathrm{mm}$
(symbols), or from $l=0$.01, 0.03, 0.1, 0.3, and 1~$\mathrm{\mu m}$ to $L=$
30~$\mathrm{mm}$ (lines), as indicated in the legend. Dashed line: free
diffusion limit. Inset: Blow-up of the log plot, focused on the small
variations induced at low pressures by changes in upper bound $L$ (at constant
lower bound $l=$0.01~$\mathrm{\mu m}$).}
\label{fig9}
\end{figure}

The effective diffusion coefficient can easily be computed from Eq.~(\ref{eq5})
using the distribution function $f(\lambda_{\mathrm{aero}})\varpropto
(\lambda_{\mathrm{aero}})^{-2+\beta}$, provided that $f$ is suitably
normalized. Since exponents $\beta$ typically range from 0.5 to 0.7 in the
aerogel samples under study, no divergence occurs at infinitely large
$\lambda$, but some cutoff must be introduced at small $\lambda$ to keep $\int
f(\lambda)d\lambda$ finite. Numerical calculations are performed to probe the
influence of both the upper and lower bounds of the mfp distribution, and
to examine the constraints experimentally set by $D_{\mathrm{eff}}$ and its
variation with $P_{3}^{\mathrm{eq}}$ in the 10~$\mathrm{mbar}-$3~$\mathrm{bar}$ range.
For simplicity, $f(\lambda_{\mathrm{aero}})$ is assumed to be zero outside the
interval $l<\lambda_{\mathrm{aero}}<L.$ Fig.~\ref{fig9} displays the results
obtained at fixed $\beta$, arbitrarily set equal to 0.5, for different values
of parameters $l$ and $L$. 
As expected, the broader the mfp distribution, the closer to a power-law
dependence is the variation of $D_{\mathrm{eff}}$ with pressure. Variations of
the upper bound $L$ lead to small changes of $D_{\mathrm{eff}}$ that can only
be noticed at low pressures. These changes are found to be quite negligible
above 10~$\mathrm{mbars}$ for $L$ greater than 1~$\mathrm{mm}.$ The lower bound
$l$ has a much stronger impact. $1/D_{\mathrm{eff}}$ increases when $l$ is
reduced, which corresponds to the aerogel hindering gas diffusion and
restricting free paths to very short distances. On the contrary if, for
instance, $l$ is raised to 1~$\mathrm{\mu m}$ the atoms move almost freely
inside the aerogel over the probed pressure range, and $D_{\mathrm{eff}}$
approaches $D_{\mathrm{gas}}$. As the exponent of the pressure dependence
exhibits no noticeable drift towards 1 at high pressure (log plot in
Fig.~\ref{fig9}), the experimental data definitely rule out $l>30$%
~$\mathrm{nm}$: mfp's on the order on a few nanometers must be encountered
in all the aerogels under study.

Taking advantage of the weak impact of the upper bound $L$ (all the more
negligible since $P_{3}^{\mathrm{eq}}<$100-$\mathrm{mbar}$ data have been discarded in
the first of all samples but aerogel A), one can actually extract from the
prefactor values $\alpha$ in Table~\ref{table1} estimates for the smallest
mfp encountered in each sample. Computing 1/$D_{\mathrm{eff}}$ as a
function of $l$ using Eq.~(\ref{eq5}) and the fit exponents $\beta$ ($L$ being
conservatively set to 3000~$\mathrm{\mu m}$), at fixed pressure $P_{3}^{\mathrm{eq}}%
=1$~$\mathrm{bar}$, adjustments to the fit prefactors $\alpha$ yield results
with quite reasonable orders of magnitude: $l=$29, 24, 22 and 24~$\mathrm{nm}$
for samples J$_{1}$, J$_{2}$, J$_{3}$, and J$_{4}$, respectively;
$l=$31~$\mathrm{nm}$ for aerogel M; $l=6.5$~$\mathrm{nm}$ only for aerogel A,
where $^{3}\mathrm{He}$ atoms experience the most restricted diffusion.

More numerical results can be found in the supplemental material provided with
this article through EPAPS \cite{EPAS}. In particular, shunt effects associated with the
presence of macroscopic aerogel-free spaces inside the experimental cells are
quantitatively investigated. Their impact is small for volume fractions below
$10\%$, and would mainly result in a downwards curve bending at low
$P_{3}^{\mathrm{eq}}$ that is not observed experimentally. Other types of broad mfp
distributions (e.g., Gaussian, bimodal, and exponential
distributions) are also easily ruled out, as computer calculations lead through
Eq.~(\ref{eq5}) to a pressure dependence that is qualitatively very
different from the observed one.

In summary, the spin diffusion measurements performed in aerogel strongly
suggest a description of this particular medium by a broad range of mfp's with power-law statistical distribution.\ The global effect of silica network
and its complex structure is best modeled with $f(\lambda_{\mathrm{aero}%
})\varpropto(\lambda_{\mathrm{aero}})^{-\nu}$, with $1.3<\nu<1.6$ depending on
the investigated aerogel sample. Data analysis indicates that these
distributions must extend at least from less than 30~$\mathrm{nm}$ to more than
hundreds of micrometers.

\subsection{Comparison with other studies}

To our best knowledge, there is no comparable quantitative study of gas
(self-)diffusion in aerogels \cite{note2}. Restricted gas diffusion has been
reported in a recent attempt to probe aerogel pore-structure with imaging and
spectroscopy NMR techniques using thermally polarized $^{129}\mathrm{Xe}$ gas
\cite{Gregory98,Gregory03}. A 3-fold decrease of the $\mathrm{Xe}$ diffusion
coefficient is observed for pressures ranging from 1 to 30~$\mathrm{bars}$, in
several small specimens of unspecified origin and nominal characteristics.
Other dense probe fluids can also be used to probe silica aerogels, e.g.,
water \cite{Gallegos88} or methanol \cite{Behr98}. But liquid phase NMR
involves much shorter relaxation times and the quantitative analysis of the
attenuation data is much less straightforward (spin-spin or spin-lattice
relaxation processes must be taken into account \cite{Behr98}). Most reported
`gas diffusion' measurements \cite{note2} are gas phase studies that actually
consist in transport measurements (permeametry), with assessment of gas flow
through virgin \cite{Stumpf92,Satoh95,Reichenauer95,Hosticka98} or partially
densified \cite{Reichenauer95,Beurroies95,Hasmy95} aerogels. In this case, a
mean pore size of the interconnected open network is usually inferred from
measured flow rates of gas through the pores using a suitable geometrical
model of the open system.

Flow measurements actually provide information about the structure of the
porous matrix in the Knudsen regime, i.e., at low enough pressure, when free
molecular diffusion dominates. The gas-flow conductance then becomes pressure
independent, and it is solely determined by the cumulative effect of
collisions on the silica walls \cite{Hasmy95,Levitz93,Hasmy98,Hasmy99}. The
standard capillary-bundle model for porous media appears to give an acceptable
description of both static or dynamic flow experiments in aerogel
\cite{Stumpf92,Satoh95,Reichenauer95,Beurroies95}. Using a tortuosity
coefficient $\tau$ to account for the pore morphology and connectivity of the
pore network, the diffusion coefficient is written as $D_{\mathrm{eff}%
}=D_{\mathrm{Kn}}/\tau$ with reference to the Knudsen diffusion coefficient
$D_{\mathrm{Kn}}$ in straight parallel, non-overlapping cylindrical pores of
infinite length. In this model $D_{\mathrm{Kn}}=v\phi/3$, where the tube
diameter $\phi$ is equal to the characteristic size of the real pore structure
under consideration (e.g., the hydraulic diameter, or mean intercept length).
The $\phi$ values derived from these molecular flow measurements range from 10
to 50~$\mathrm{nm}$, in agreement with the hydraulic diameters computed from
standard stereology: $\phi=4\epsilon/\rho\Sigma$ ($\epsilon$: porosity, $\rho
$: density, and $\Sigma$: specific area) or the mfp's obtained from SAS
measurements \cite{Beurroies95,Reichenauer95}. And the effective diffusion
lengths are 1.4 to 3 times larger than for straight line motion, due to
tortuosity \cite{Satoh95,Reichenauer95}.

Extending the capillary-bundle model to the aerogels used in the present work,
in the Knudsen regime one would expect diffusion coefficients on the order of
0.6~$\mathrm{cm}^{2}\mathrm{/s}$ for samples of both type A and J, since the
hydraulic diameters computed from their commercially specified specific areas
and densities are quite comparable for these two ($\phi$=127 and
116~$\mathrm{nm}$, respectively). The large differences in $D_{\mathrm{eff}}$
experimentally observed by NMR at all pressures, in the two types of aerogel,
suggest that either (1) the effective pore size derived from the
capillary-bundle model suited for the analysis of flow measurements is not
relevant to characterize the influence of the complex ramified silica network
on pure self-diffusion in the gas, or (2) the structure of the aerogel matrix
in the high porosity samples under study is very different from that of most
aerogels used in the above-quoted references.

In several experiments, the Knudsen regime is reported to be reached for
average gas pressures typically below 1~$\mathrm{bar}$ for aerogel densities
ranging from 0.1 to 2.2~$\mathrm{g/cm}^{3}$ \cite{Reichenauer95,Beurroies95}.
However, a clear transition from molecular to viscous flow has once been
observed around 150~$\mathrm{mbars}$ in a 0.28-$\mathrm{g/cm}^{3}$ sample (yet,
the authors computed an effective pore radius on the order of 19~$\mathrm{\mu
m},$ using a modified capillary-bundle model) \cite{Stumpf92}. Moreover, a
similar observation can be found in one recent systematic study devoted to
samples from an aerogel monolith grown with comparably high porosity (density
0.06~$\mathrm{g/cm}^{3}$; area $800\pm30~\mathrm{m}^{2}\mathrm{/g})$
\cite{Hosticka98}. Interestingly enough, the conclusion drawn from flow
measurements performed in that study is that ``the absence of any clear break
(in the transition from diffusive to viscous flow) supports the assumption of
a continuous distribution of pores sizes within the range of mean free paths
in this experiment''. The experiment was actually performed with $\mathrm{He}$
mfp's ranging from 90 to 3.5~$\mathrm{\mu m}$ (measurements
were also repeated with $\mathrm{N}_{2}$ gas, for mfp's ranging between
30 and 1~$\mathrm{\mu m}$, and a similar behavior has been
observed). Furthermore, this analysis correlated with an intriguing
``significant pore volume (detected) throughout the measuring range of the
instrument from 2 to 200~$\mathrm{nm}$'' (by nitrogen desorption isotherm
technique) \cite{Hosticka98}. Hence, the upper and lower boundaries of the
continuous pore size distribution assumed for the high porosity aerogel
specimens investigated in Ref. \cite{Hosticka98} are quite comparable to those
derived in the present NMR\ work for $^{3}\mathrm{He}$ (self-)diffusion
mfp's, using the power-law distribution model presented in
Section~\ref{section5.2}.

\section{Conclusion and prospects}

In summary, laser optical pumping has been combined with low field NMR to
nondestructively probe high porosity aerogel samples, by spin-echo diffusion
measurements with hyperpolarized $^{3}$\textrm{He }gas. Analysis of the
diffusion data has been based on a statistical description of the gas motion
inside the solid matrix. Restricted diffusion has been related to $^{3}%
$\textrm{He} mfp changes induced by interactions between the probe atoms
and the sparsely distributed silica strands, a gas kinetics approach
particularly suited to the widely open aerogels under study. The experimental
findings are satisfactorily accounted for, assuming broad mfp's
distributions inside the aerogels with ranges and power-law probability weights that are
determined from the pressure dependence of the measured effective diffusion coefficients.

The conclusions drawn from our study support the idea that these high porosity
aerogels are not well-defined systems. It may hence contribute to cast doubt
on simple models and interpretations of low-temperature physics experiments,
and fosters the growing concern that they may not be ideal host media when
high control of induced confinement and disorder is required. Moreover, some
aspects of the methodology and experimental techniques that have been used may
be directly applied, or extended, to a variety of other porous systems, as follows.

In our work, a buffer gas\ ($\mathrm{N}_{2}$) has been used to slow down
$^{3}\mathrm{He}$ diffusion in the aerogel void spaces, and to achieve a
significant increase of the pressure measurement range. This is applicable in
all systems where the gas pressure cannot be increased \textit{ad libitum}. For
instance, diffusion imaging of inhaled hyperpolarized $^{3}\mathrm{He}$ gas is
currently used to observe in human lungs spatial variations of alveolar size
related to physiology \cite{Bidinosti03, Bidinosti04}, posture
\cite{Fichele04a}, respiratory conditions \cite{Salerno02}, or disease
\cite{Saam00,Salerno02,Swift05}. Operation at atmospheric pressure being
mandatory \textit{in vivo}, mixing with biologically inert heavy buffer gases may be
advantageously used to enhance contrast for such studies (e.g., $\mathrm{SF}%
_{6}$ decreases the $^{3}\mathrm{He}$ diffusion coefficient by a factor up to
4 at 1 bar, depending on dilution) but also, more generally, to reduce
diffusion-weighted attenuation by encoding gradients for improved SNR
sensitivity and image quality \cite{Acosta04}.

Using hyperpolarized $^{3}\mathrm{He}$, we have been able to probe our aerogel
samples over a very broad range of diffusion lengths with the same
measurement\ technique. This provides a powerful alternative to standard NMR
approaches, often limited by technical aspects (e.g., acquisition rates,
gradient amplitudes or ramping times), provided that the pressure can be
directly varied by several orders of magnitude in the porous medium.
Application to denser silica aerogels is straightforward, but may be considered
as well for other rigid porous systems allowing fast gas diffusion. For
instance, diffusion measurements could be considerably extended \textit{in vitro},
using fixed lungs with preserved air spaces \cite{Acosta04}, to bridge the gap
between conflicting \textit{in vivo} lung investigations performed on time scales
ranging from a few milliseconds to tens of seconds (using gradient echo
\cite{Saam00,Salerno02}, spin echo \cite{Durand02} and spin tagging
\cite{Owers03,Woods04} techniques), and to clear up\ the link between
$^{3}\mathrm{He}$ NMR data and airways structural characteristics
\cite{Chen99,Yablonskiy02,Fichele04b}.

Laser polarization of the probe gas actually yields high SNRs to perform
highly accurate NMR measurements in a variety of model systems (single pores
\cite{Hayden04,McGregor90}, rocks \cite{Mair99} or packed glass beads
\cite{Mair99,Mair02}). The technique yet requires that a sufficiently large
nuclear polarization is preserved inside the porous medium. This potentially
limits the type of materials that can be probed, at least in the static mode
operated here. However, for materials with moderately low specific relaxation
times, continuous flow techniques can be used to maintain a suitable
stationary magnetization, through replenishment by freshly polarized gas. Most
demonstrations involve hyperpolarized xenon gas so far \cite{Mair02,Wang04},
but flow measurements are also very easy to perform with the compact
$^{3}\mathrm{He}$ gas production system described in Section~\ref{section2.3}.

Finally, a specific interest of operation at low magnetic field may be
emphasized. Interpretation of NMR diffusion data is known to crucially depend
on adequate background gradient compensation, but also on the predominance of
applied gradients over internal ones. In our work, local-field inhomogeneities
due to the aerogel matrix were irrelevant for the analysis of the
measurements. But NMR studies with hyperpolarized gas are highly sensitive,
and they may equally reveal weak bulk magnetic susceptibility effects (e.g.,
the influence of glass containers at high field \cite{Saam96}) or artifacts
generated by dominant susceptibility-induced field gradients\ (e.g., in
ventilation imaging at 1.5 $~\mathrm{T}$ \cite{Salerno01,vanBeek04}).
Operation at low or ultralow field can be very helpful to separate the
effects of applied and local gradients for improved quantitative understanding
in complex porous systems, as recently demonstrated in lungs
\cite{Durand02,Bidinosti03,Bidinosti04}.

\begin{acknowledgments}
Sample M has been kindly prepared for us by Norbert Mulders in Moses Chan's
laboratory at Pennstate University. Similar custom-made aerogels are used for 
low-temperature experiments at
UPenn, Cornell and Northwestern Universities, and several
other places.
Samples A and J$_{1}-$J$_{4}$ have been provided by Georg Eska,
Bayreuth University.
\end{acknowledgments}

\section*{Supplemental material: EPAS Document E-JCPSA6-123-034528 \cite{EPAS}}

All the notations used in this supplemental material are those introduced in
the related article.

The first section investigates the effects of macroscopic void (aerogel-free) spaces present
in the experimental volumes on the NMR diffusion measurements with
hyperpolarized $^{3}\mathrm{He}$ gas. The probe atoms are strongly confined
inside the aerogel, but diffuse freely (hence, much faster) in these void
spaces. If the contribution of such diffusion ``shunts'' to the spin-echo
attenuation is significant, the global effective coefficient $D_{\mathrm{eff}%
}$ extracted from the NMR data must differ from that of the bulk aerogel.
Shunt effects are quantitatively assessed, and discussed as a function of the
volume fraction of open void spaces.

The second section provides examples of computed effective diffusion
coefficients in bulk aerogel, for a few kinds of broad mfp distributions.
Results are successively presented for Gaussian, bimodal, and exponential
distributions. They contrast strongly with those obtained with the power-law
distributions used in the article, that satisfactorily describe the
experimental diffusion data. 

\subsection{Shunt effects}

Experimentally, the void spaces open to $^{3}\mathrm{He}$ diffusion may be
large cracks that have developed inside the aerogel samples, or gaps between
the bulk material and the container walls for aerogel samples that do not fill
up the whole cell volume.

\begin{figure}[h]
\includegraphics[clip,width=3.15 in]{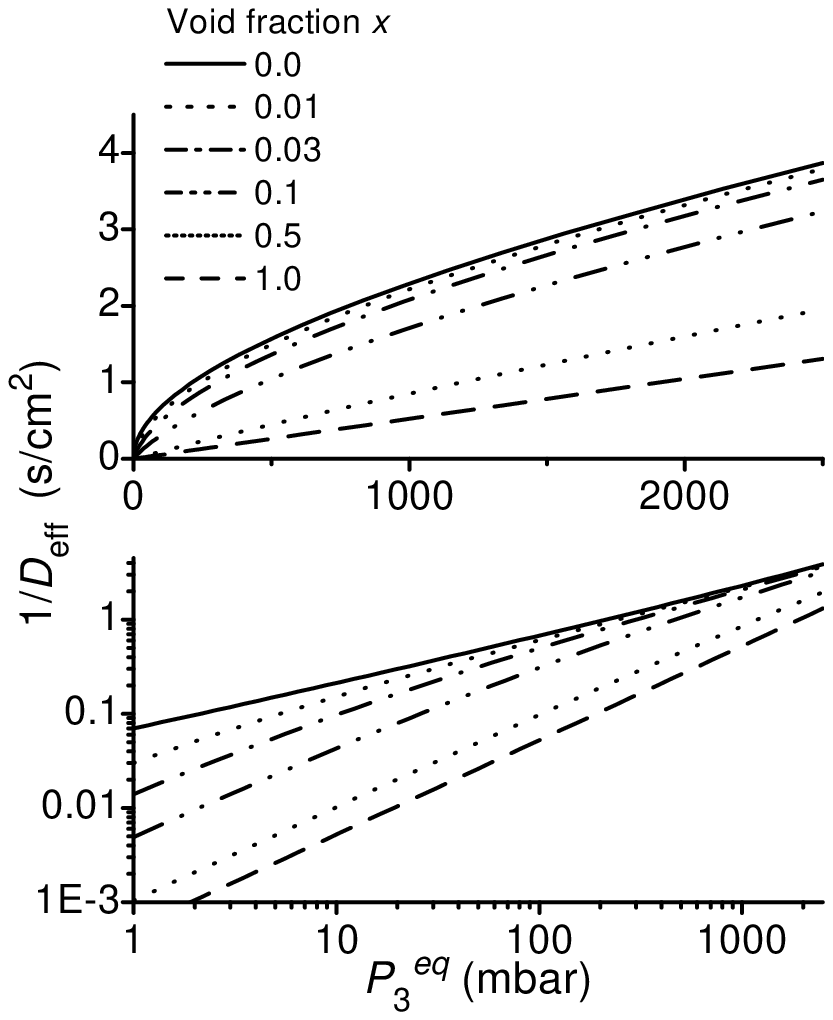}
\caption{Computed pressure
dependence of $1/D_{\mathrm{eff}}$ as a function of the void fraction $x$
[Eq.~(S-1)], for a mfp distribution in aerogel $f(\lambda_{\mathrm{aero}%
})\varpropto(\lambda_{\mathrm{aero}})^{\beta-2}$ with $\beta=0.5$, ranging
from $l=$10~$\mathrm{nm}$ to $L=$30~$\mathrm{mm}$.}
\label{fig10}%
\end{figure}

A finite fraction $x$ of the sample volume where mfp $\lambda
_{\mathrm{aero}}$ gets infinite is introduced to describe such a macroscopic
``aerogel-free'' region of space, where the atoms are free to move. As in
Section V-B of the related article, fast exchange is assumed between this region and
the bulk aerogel, leading to an effective diffusion coefficient that can be
written as:
\begin{equation}
D_{\mathrm{eff}}=(1-x)\int\frac{f(\lambda_{\mathrm{aero}})\mathcal{D}%
_{3}D_{\mathrm{aero}}(\lambda_{\mathrm{aero}})d\lambda_{\mathrm{aero}}%
}{\mathcal{D}_{3}+D_{\mathrm{aero}}(\lambda_{\mathrm{aero}})P_{3}^{\mathrm{eq}}}%
+x\frac{\mathcal{D}_{3}}{P_{3}^{\mathrm{eq}}}\text{.}\tag*{(S-1)}\label{S-1}%
\end{equation}
Fig.~\ref{fig10} shows the variations of $1/D_{\mathrm{eff}}$ with the void
fraction $x$, for a fixed mfp distribution in aerogel (ranging from
10 to 30~$\mathrm{mm}$, with $\beta=$0.5).
For all pressures, trivially, $D_{\mathrm{eff}}$ increases towards
$D_{\mathrm{gas}}$ as $x$ increases. However, large global changes of
$1/D_{\mathrm{eff}}$ (i.e., of spin-echo decay times) occur only for $x>10\%.$
The changes induced by the void fraction are relatively larger at lower
pressures. At small $x$, the log plot indicates that they translate into a
significant downwards bending of the curves at low $P_{3}^{\mathrm{eq}}$ that is not
observed experimentally. At intermediate $x$, the pressure dependence is
strongly altered over the whole pressure range, and cannot be satisfactorily
described by a single power-law variation.

In the custom-made aerogel M (grown-in sample, attached to the glass walls),
the void fraction corresponding to the visually observed cracks is very likely
to be extremely small. Since no significant departure from the power-law fit
is observed in the data down to $P_{3}^{\mathrm{eq}}=10$~$\mathrm{mbars}$, their actual
contribution to $D_{\mathrm{eff}}$ is negligible.

Void spaces are larger in the commercial aerogel samples. Some have shape
imperfections (e.g., small pieces of material missing on the corners, and
irregular edge surfaces) that leave macroscopic open spaces with dimensions
below 1~$\mathrm{mm}$. All samples, irreversibly shrunken after a few pressure
cycles, no longer occupy the whole volume of the containers. Conservative
estimates for the volume fraction $x$ hardly reach 10$\%-$20$\%$. The largest
void fraction must be that of sample J$_{1}$ (crushed sample with trapezoidal
shape, already not matching its container in thickness). But, the exponent
associated to the pressure dependence measured in aerogel J$_{1}$ is actually
smaller than those obtained in aerogels J$_{2}$ and J$_{4}$, and not larger.

All this supports our belief that the measured features can trustfully be
assigned to be those of bulk aerogel.

\subsection{Diffusion coefficients computed for various mfp distributions.}

The expected pressure dependence of $1/D_{\mathrm{eff}}$ can easily be
computed for any type of normalized mfp distribution function using Eq.~(10)
of the related article. The expression of the effective diffusion coefficient
may be slightly modified by simple algebra (using $P_{\mathrm{aero}}%
(\lambda_{\mathrm{aero}})=3\mathcal{D}_{3}/v\lambda_{\mathrm{aero}}$), to let
the rational function of pressure and mfp involved in the integrand appear
more explicitly:
\begin{equation}
D_{\mathrm{eff}}=\int\frac{f(\lambda_{\mathrm{aero}})\mathcal{D}_{3}%
d\lambda_{\mathrm{aero}}}{P_{\mathrm{aero}}(\lambda_{\mathrm{aero}}%
)+P_{3}^{\mathrm{eq}}}=\frac{\mathcal{D}_{3}}{P_{1}}\int\frac{f(\lambda)\;\lambda
\;d\lambda}{\lambda_{1}+\lambda\;(P_{3}^{\mathrm{eq}}/P_{1})}\tag*{(S-2)}\label{S-2}%
\end{equation}
where $P_{1}$ is an arbitrary reference value for pressure, and $\lambda
_{1}=3\mathcal{D}_{3}/vP_{1}$ the corresponding mfp for free diffusion
(e.g., $\lambda_{1}=399.2~\mathrm{\mu m}$ at $293~\mathrm{K}$ for
$P_{1}=1~\mathrm{mbar}$). Some examples are briefly discussed below (Gaussian,
bimodal and exponential distributions), to show the qualitative differences
between these computed pressure dependences and the experimentally observed
features. 

For convenience, all computed values of $1/D_{\mathrm{eff}}$ are discussed in
terms of $P_{\mathrm{eff}}=\mathcal{D}_{3}/D_{\mathrm{eff}}$ (the pressure\ of
freely diffusing $^{3}\mathrm{He}$ gas required to yield a diffusion
coefficient value equal to $D_{\mathrm{eff}}$). Hereafter, the probing gas
pressure is also simply designated by $P_{3}$ (equal to the actual pressure
for pure $^{3}\mathrm{He}$ gas, or to the equivalent pressure $P_{3}^{\mathrm{eq}}$ for
a gas mixture as described in the text for the $^{3}\mathrm{He-N}_{2}$
mixtures used in the experiments).

\subsubsection{Gaussian mfp distributions\label{sectionA}}

\begin{figure}[h]
\includegraphics[clip,width=3.15 in]{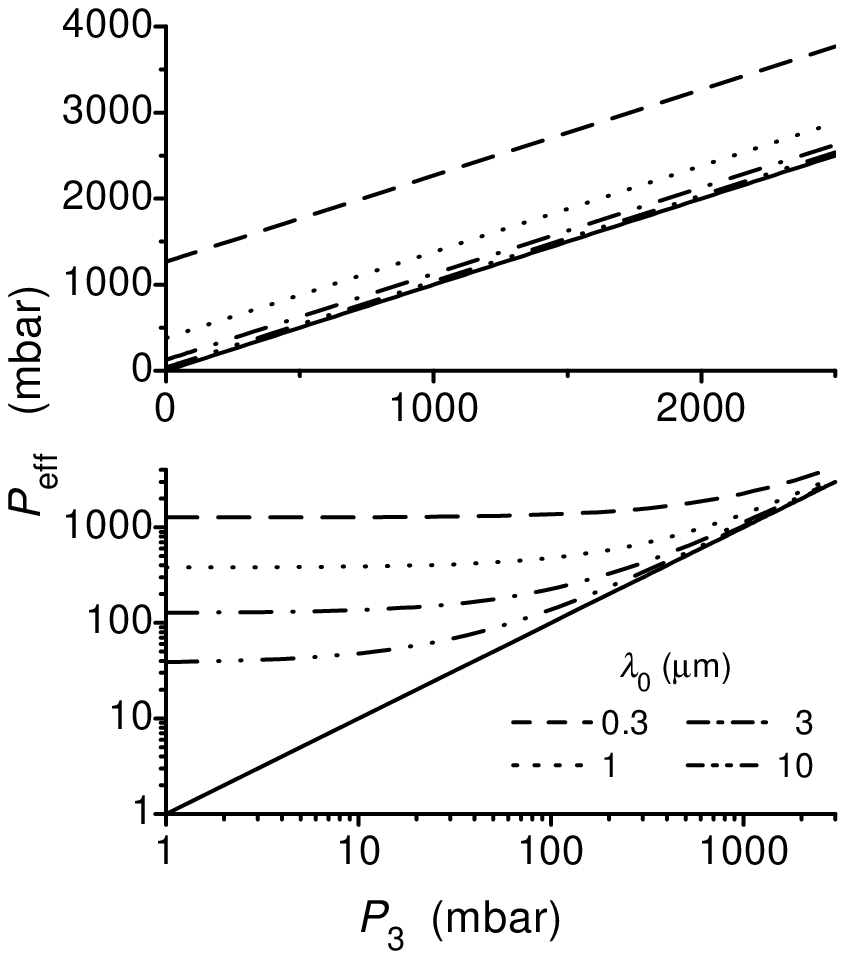}
\caption{Broken lines: Pressure dependence of
$P_{\mathrm{eff}}=\mathcal{D}_{3}/D_{\mathrm{eff}}$
[Eq.~(S-2), with $f(\lambda)\varpropto\exp[-(\lambda-\lambda_{\mathrm{0}}%
)^{2}/\delta\lambda^{2}]$], computed for various mean values $\lambda
_{\mathrm{0}}$ (see legend) and $\delta\lambda=0.01~\mathrm{\mu m}$. Solid
line: free gas data ($P_{\mathrm{eff}}=P_{3}$).}
\label{fig11}
\end{figure}

\begin{figure}[h]
\includegraphics[clip,width=3.15 in]{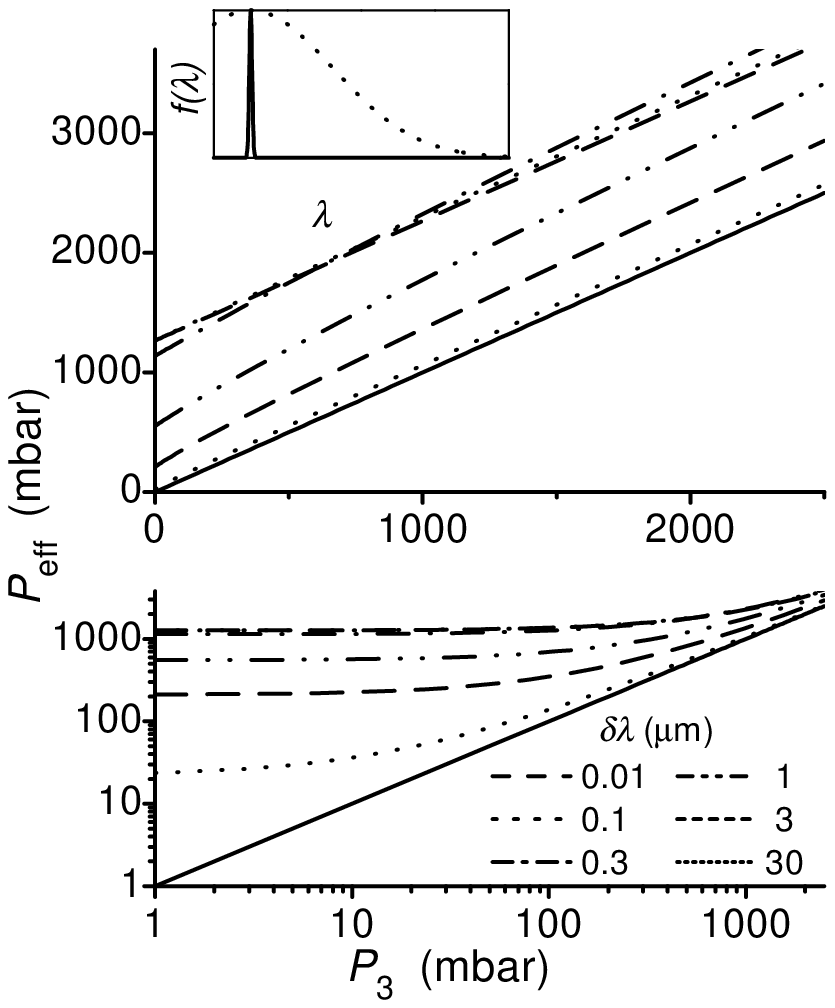}
\caption{Broken lines: Pressure dependence of
$P_{\mathrm{eff}}=\mathcal{D}_{3}/D_{\mathrm{eff}}$
for Gaussian mfp distributions in aerogel [Eq.~(S-2), with $f(\lambda
)\varpropto\exp[-(\lambda-\lambda_{\mathrm{0}})^{2}/\delta\lambda^{2}]$],
computed for various widths $\delta\lambda$ (see legend) and constant mean
value $\lambda_{\mathrm{0}}=0.3~\mathrm{\mu m}$. Solid line: free gas
diffusion ($P_{\mathrm{eff}}=P_{3}$). Inset: Illustration of two
extreme cases: broad (dashed line) and narrow (solid line) Gaussian mfp distributions.
For the broad one, truncation at $l=0$ leads to an average mfp that is
significantly higher (hence, to less restricted atomic diffusion).}
\label{fig12}
\end{figure}

Figs.~\ref{fig11} and \ref{fig12} display the results obtained for Gaussian
mfp distributions characterized by mean values $\lambda_{\mathrm{0}}$ and
widths $\delta\lambda.$

A narrow Gaussian distribution leads to a variation of $P_{\mathrm{eff}}$ that
is very similar to that described in Section~V-A of the related article for a
uniform medium with a single characteristic mfp $\lambda_{\mathrm{aero}}$
equal to $\lambda_{\mathrm{0}}.$ For instance, selected data from computations
performed for various mean values $\lambda_{\mathrm{0}}$ and constant width
$\delta\lambda$ are presented in Fig.~\ref{fig11}.
The linear plot in Fig.~\ref{fig11} shows that, for each
$\lambda_{\mathrm{0}}$, $P_{\mathrm{eff}}$ actually scales linearly with
$P_{3}$ with an offset value that increases as $\lambda_{\mathrm{0}}$
decreases. With $\delta\lambda=0.01~\mathrm{\mu m}$, over the probed pressure
range, all curves can be fit by straight lines with slopes equal to 1 and
offset values $P_{\mathrm{eff}}(0)$ $=3\mathcal{D}_{3}/v\lambda_{\mathrm{0}}$
with an accuracy better than 50 ppm. The log plot in Fig.~\ref{fig11}
therefore exhibits, for each value of $\lambda_{\mathrm{0}}$, a crossover from
the free gas behavior (solid line) at high pressure, to a plateau at low
pressure (where the mfp for free diffusion becomes so large that the
effective mfp is just that imposed by the aerogel, $\lambda_{\mathrm{0}}$).
This crossover occurs when the mfp $\lambda_{^{3}\mathrm{He}}$ for gas
collisions becomes comparable to $\lambda_{\mathrm{0}}$ (i.e., around
$P_{3}=3\mathcal{D}_{3}/v\lambda_{\mathrm{0}}$). The crossover region
typically extends over one decade in pressure.

The influence of the width $\delta\lambda$ is depicted in Fig.~\ref{fig12},
where $\lambda_{\mathrm{0}}$ is kept constant and $\delta\lambda$ is varied
($\lambda_{\mathrm{0}}=0.3~\mathrm{\mu m}$; $\delta\lambda=0.01$ to
$30~\mathrm{\mu m}$).
The impact of the Gaussian width is quite small as long as
$\delta\lambda\leqslant$ $\lambda_{\mathrm{0}}$, and corresponds to a slight
change of the effective slope (linear plot in Fig.~\ref{fig12}: dash, dot, and
dash-dot lines). A significant change occurs when $\delta\lambda$ becomes
comparable to $\lambda_{\mathrm{0}}$, due to truncation at small
$\delta\lambda$. This situation is qualitatively depicted by the inset in
Fig.~\ref{fig12} that displays Gaussian distribution profiles for
$\delta\lambda\ll$ $\lambda_{\mathrm{0}}$ (solid line) and for $\delta
\lambda>$ $\lambda$ (dashed line). For a further increase of the Gaussian
width, the impact becomes more significant since much larger mfp's are
reached (looser spatial confinement of the atoms), and a strong offset
decrease is observed for $\delta\lambda\gg\lambda_{\mathrm{0}}$ (linear plot
in Fig.~\ref{fig12}: dash-dot-dot, short dash, and short dot lines). For
instance, in spite of the short mfp mean value selected here ($\lambda
_{\mathrm{0}}=0.3~\mathrm{\mu m}$), the free-diffusion limit would nearly be
recovered over the experimentally probed pressure range (i.e., $P_{3}%
>10~\mathrm{mbars}$, and hence $\lambda_{\mathrm{gas}}<40~\mathrm{\mu m}$) for
$\delta\lambda$ on the order of $30$~$\mathrm{\mu m}$ (log plot in
Fig.~\ref{fig12}).

As a conclusion, whatever the choice of numerical parameters, a Gaussian
distribution fails to provide a suitable pressure dependence to account for
the experimental measurements.

\subsubsection{Bimodal mfp distributions}

\begin{figure}[h]
\includegraphics[clip,width=3.15 in]{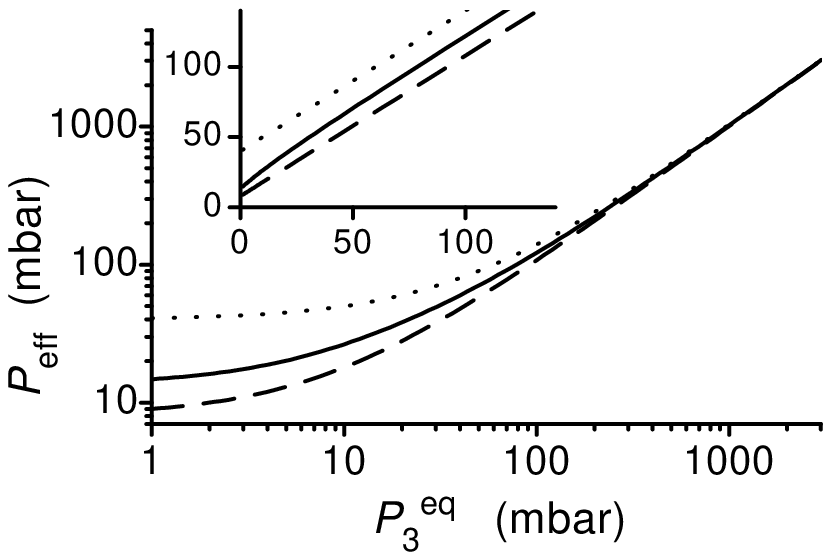}
\caption{Pressure dependence of $P_{\mathrm{eff}}$ obtained for a bimodal mfp distribution
(solid line) that combines two narrow Gaussian distributions : $\lambda
_{\mathrm{0}}=50~\mathrm{\mu m,}$ $\delta\lambda=0.1~\mathrm{\mu m}$
(distribution $f^{(1)}$, dashed line), and $\lambda_{\mathrm{0}}%
=10~\mathrm{\mu m,}$ $\delta\lambda=0.1~\mathrm{\mu m}$ (dotted line), in log
plot. Inset: linear plot, for the same distributions.}
\label{fig13}
\end{figure}

\begin{figure}[h]
\includegraphics[clip,width=3.15 in]{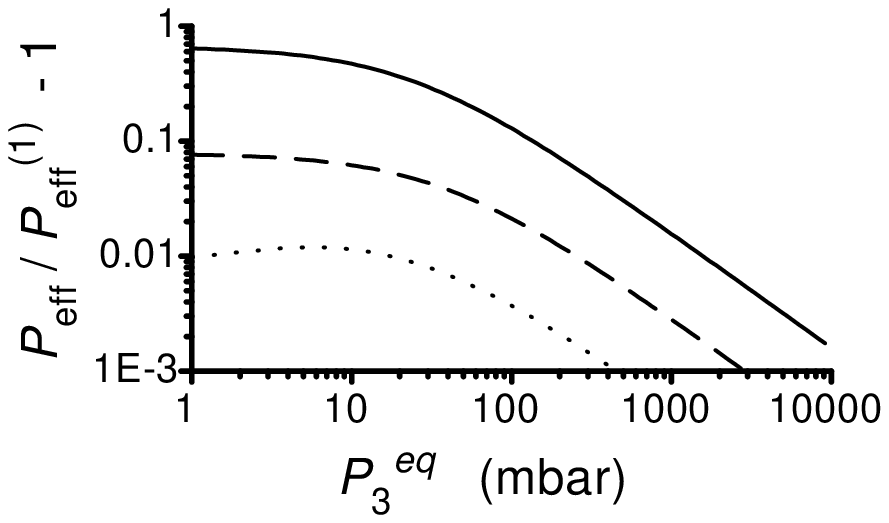}
\caption{Pressure dependence of the relative changes in $P_{\mathrm{eff}}$ for 
bimodal mfp distributions obtained by combination of Gaussian
distributions $f^{(1)}$ centered around
$\lambda_{\mathrm{0}}=50~\mathrm{\mu m}$\textrm{\ } with
various widths $\delta\lambda$ (providing the reference values
$P_{\mathrm{eff}}^{(1)}$), and a fixed Gaussian distribution (mean:
$10~\mathrm{\mu m,}$\textrm{\ }width: $0.1~\mathrm{\mu m}$). Solid line:
$\delta\lambda=0.1~\mathrm{\mu m}$ (value identical to that used in
Fig.~\ref{fig13}). Dashed line: $\delta\lambda=1~\mathrm{\mu m}$. Dotted
line: $\delta\lambda=10~\mathrm{\mu m}$.}
\label{fig14}
\end{figure}

Figs.~\ref{fig13} and \ref{fig14} illustrate the expected behaviors for
bimodal distributions that are sums of two Gaussian functions with identical amplitudes.

In Fig.~\ref{fig13}, a narrow Gaussian distribution $f^{(1)}$ ($\lambda
_{\mathrm{0}}=$ $50~\mathrm{\mu m}$, $\delta\lambda=0.1~\mathrm{\mu m)}$ is
combined with another narrow Gaussian distribution (mean: $10$~$\mathrm{\mu
m}$, width: $0.1~\mathrm{\mu m}$).
As expected, the dominant contribution arises from the
distribution that provides access to the largest mfp's. The relative difference
between the bimodal distribution and the $f^{(1)}$ Gaussian distribution is
noticeable at low pressure. The combination with the other Gaussian results in
a significant offset increase ($70\%$ change), and a slight curve distortion
below $P_{3}=100~\mathrm{mbars}$ (best seen in the inset). The corresponding
relative change in $P_{\mathrm{eff}}$ values is plotted against gas pressure
in Fig.~\ref{fig14} (solid line).
It becomes negligible as soon as the free gas mfp gets smaller
than the lowest bound of the mfp distribution in aerogel.

Fig.~\ref{fig14} also displays results obtained for different values of the
width $\delta\lambda$ of the $f^{(1)}$ Gaussian distribution, that remains
centered around $\lambda_{\mathrm{0}}=50~\mathrm{\mu m.}$ The relative impacts
of the combination with the other Gaussian distribution can be checked to
become smaller as the width $\delta\lambda$ is increased, due to
the reduced relative weight of this additional contribution (Fig.~\ref{fig14},
broken lines).

Whatever the combination of Gaussian distributions performed, bimodal
distributions can absolutely not be adjusted to match the power-law pressure
dependence that is experimentally observed.

\subsubsection{Exponential mfp distributions}

\begin{figure}[h]
\includegraphics[clip,width=3.15 in]{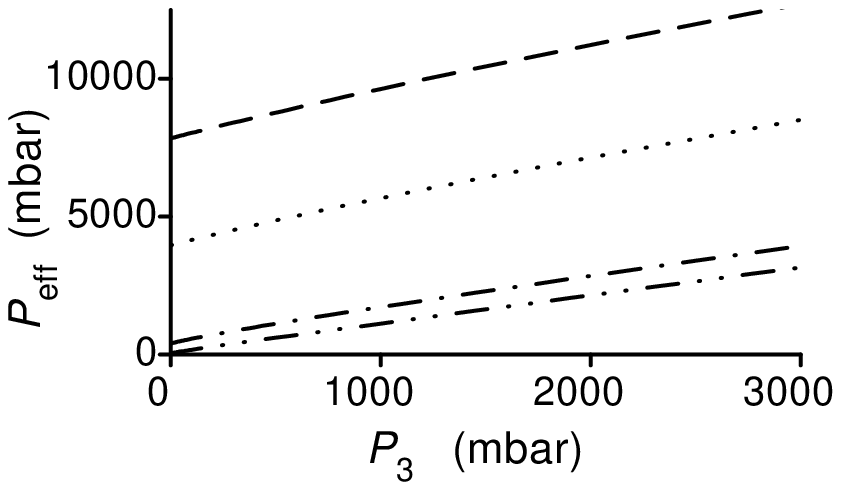}
\caption{Pressure dependence of $P_{\mathrm{eff}}$ for exponential mfp distributions [Eq.~(S-2),
with $f(\lambda)\varpropto\exp(-\lambda/\lambda_{\mathrm{c}})$] for various
characteristic mfp's $\lambda_{\mathrm{c}}$: $0.05~\mathrm{\mu m}$ (dashed
line), $0.1~\mathrm{\mu m}$ (dotted line), $1~\mathrm{\mu m}$ (dash-dot line), and
$10~\mathrm{\mu m}$ (dash-dot-dot line).}
\label{fig15}
\end{figure}

Fig. \ref{fig15} displays the pressure dependences typically obtained for
exponential distributions with characteristic mfp's $\lambda_{\mathrm{c}}$
ranging from $0.05$ to $1~\mathrm{\mu m}$ (identified by distinct line styles
in the graph).
$P_{\mathrm{eff}}$ remains finite at very low pressures and
approaches $P_{\mathrm{eff}}(0)=3\mathcal{D}_{3}/v\lambda_{\mathrm{c}}$, the
expected value for an unbounded distribution. $P_{\mathrm{eff}}$ also
increases with pressure, but not linearly. The derivative 
$dP_{\mathrm{eff}}/dP_{3}$ smoothly varies
between two asymptotic values: 2, at null pressure, and 1, at infinite
pressure. For each $P_{\mathrm{eff}}$ curve, the change of local slope
essentially occurs over the pressure range where the mfp for free diffusion
is larger than (or equal to) the characteristic scale $\lambda_{\mathrm{c}}$
introduced by the aerogel. Beyond this limit, the gas motion is hardly limited
by the solid matrix. This nearly-free diffusion regime (linear pressure
dependence with slope 1) is reached for $P_{3}\gg40~\mathrm{mbars}$ at
$\lambda_{\mathrm{c}}=10~\mathrm{\mu m}$, for instance. At $\lambda
_{\mathrm{c}}=0.05~\mathrm{\mu m}$, the probed pressure range must be extended
to $P_{3}\gg8~\mathrm{bars}$ to observe it (data not shown in Fig.\ref{fig15}).

\begin{figure}[h]
\includegraphics[clip,width=3.15 in]{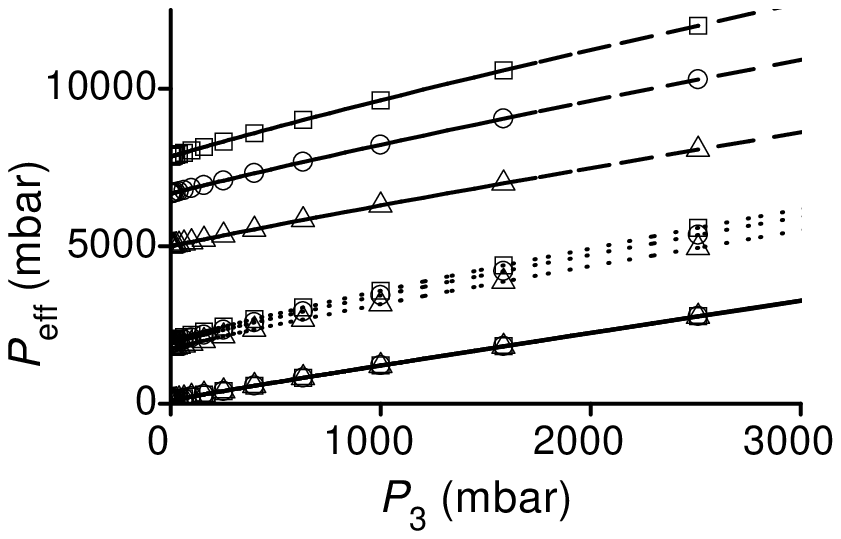}
\caption{Pressure
dependence of $P_{\mathrm{eff}}$ for truncated exponential mfp
distributions. Computations are performed for three values of the
characteristic mfp $\lambda_{\mathrm{c}}$: $0.05~\mathrm{\mu m}$ (dashed
lines), $0.2~\mathrm{\mu m}$ (dotted lines), and $5~\mathrm{\mu m}$ (solid lines).
For each $\lambda_{\mathrm{c}}$ value, three different lower bounds $l$ are
set to the mfp distribution: $0.001~\mathrm{\mu m}$ (squares),
$0.01~\mathrm{\mu m}$ (circles), and $0.03~\mathrm{\mu m}$ (triangles).}
\label{fig16}
\end{figure}

When a lower bound $l$ is arbitrarily set to the exponential mfp
distribution, the results for $P_{\mathrm{eff}}$ are strongly altered
(Fig.\ref{fig16}).
Since all mfp's smaller than $l$ are removed from the
distribution, a global reduction of effective pressure may be expected (lesser
gas confinement). Quantitatively, the new value expected at zero pressure is
$P_{\mathrm{eff}}(0)=3\mathcal{D}_{3}/v(\lambda_{\mathrm{c}}+l)$, and the
relative impact of the truncation rapidly increases as $l/\lambda_{\mathrm{c}%
}$ increases. The truncation also induces a change of the $P_{\mathrm{eff}}$
increase rate over the whole pressure range, related to the modification of
the probability distribution. This effect depends on the lower bound values
$l$, as shown in Fig.\ref{fig16} for three exponential distributions (here
also, line styles vary with characteristic\ mfp values $\lambda
_{\mathrm{c}}$). For instance, detailed comparison of the results obtained
with lower bounds $l=0.01~\mathrm{\mu m}$ (circles) and $l=0.03~\mathrm{\mu
m}$ (triangles), with reference to those obtained with $l=0.001~\mathrm{\mu
m}$ (squares) in Fig.\ref{fig16} yields the following results: 1/ The relative
$P_{\mathrm{eff}}$ changes computed at fixed pressure, for $P_{3}%
=4.8~\mathrm{bars}$, amount to $30\%$, $10\%$ and $0.5\%$ for $\lambda
_{\mathrm{c}}=0.05$, $0.2$, and $5~\mathrm{\mu m}$ (respectively) with lower
bound $l=0.01~\mathrm{\mu m}$. They drop to $14\%$, $4\%$ and $0.2\%$ for
$\lambda_{\mathrm{c}}=0.05$, $0.2$, and $5~\mathrm{\mu m}$ (respectively) with
lower bound $l=0.03~\mathrm{\mu m}$. 2/ These $P_{\mathrm{eff}}$ changes
decrease slowly, but steadily, as the gas pressure increases. From
$P_{3}=1~\mathrm{mbar}$ to $P_{3}=10~\mathrm{bars}$, the exhibited variation is
equal to $27\%$ with lower bound $l=0.01~\mathrm{\mu m}$, and drops to $14\%$
with lower bound $l=0.03~\mathrm{\mu m}$, for the three $\lambda_{\mathrm{c}}$ values.

\begin{figure}[h]
\includegraphics[clip,width=3.15 in]{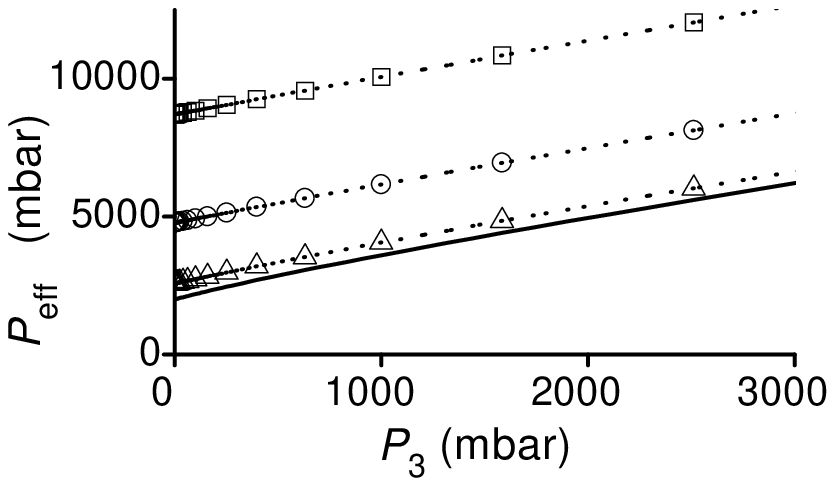}
\caption{Pressure dependence of $P_{\mathrm{eff}}$ for an exponential mfp distribution,
with (broken lines) or without (solid line) truncation at large mfp's. The
characteristic mfp of the distribution is $\lambda_{\mathrm{c}%
}=0.2~\mathrm{\mu m}$. Data are computed for three upper bound $L$ values:
$0.1~\mathrm{\mu m}$ (squares), $0.2~\mathrm{\mu m}$ (circles), and
$0.5~\mathrm{\mu m}$ (triangles).}
\label{fig17}
\end{figure}

The distribution may also be truncated at large mfp's to probe the
sensitivity of the diffusion coefficient to such a limit. The influence of the
upper bound $L$ remains extremely limited due to the exponentially vanishing
statistical weight of the large aerogel mfp's, as long as the selected $L$
value exceeds that of the characteristic mfp $\lambda_{\mathrm{c}}$.
Conversely, it is important for upper bounds $L$ comparable to $\lambda
_{\mathrm{c}}$. The effective pressure at zero gas pressure depends, quite
naturally, on the exponential factor $r=\exp(-L/\lambda_{\mathrm{c}})$, and it
is expected to vary as $P_{\mathrm{eff}}(0)=(3\mathcal{D}_{3}/v\lambda
_{\mathrm{c}})\times\lbrack1+r\ln(r)\ /(1-r)]^{-1}$. But $P_{\mathrm{eff}}$
changes must noticeably vary with gas pressure, being much larger at low
pressures (restriction of atomic diffusive motion over large length scales has
been lifted) than at high pressures (atom-atom scattering already sets a small
mfp limit). In Fig.\ref{fig17}, dotted lines show, for instance, the
results obtained for truncated exponential distributions with $\lambda
_{\mathrm{c}}=0.5~\mathrm{\mu m}$, and comparable upper mfp bounds $L$.
Symbols are used to specify the selected $L$ values, that vary
from $0.1$ to $0.5~\mathrm{\mu m}$ (see caption). The $P_{3}=1~\mathrm{mbar}$
data differ from the zero pressure expected values $P_{\mathrm{eff}}(0)$ by
less than $0.2\%$, for the three $L$ bounds. With reference to the unbounded
distribution data (Fig.\ref{fig17}, solid line), the $P_{\mathrm{eff}}$
increases due to the truncation vary from $4.35$ (respectively: $2.39$, and
$1.29$) at $P_{3}=1~\mathrm{mbar,}$ to $1.48$ (respectively: $1.19$, and
$1.04$) at $P_{3}=10~\mathrm{bars}$, for $L=0.1~\mathrm{\mu m}$ (respectively:
$L=0.2~\mathrm{\mu m}$, and $L=0.5~\mathrm{\mu m}$).

With or without truncation, exponential mfp distributions fail to provide
effective diffusion coefficients that behave like the experimental ones.

In summary, for the three types of mfp distributions presented here, the
computed pressure dependences exhibit features that do not correspond to the
observed ones. A satisfactory description of the experimental diffusion data
can only be obtained with the phenomenological power-law distributions introduced
and characterized in the related article (Section V-B).


\begin{thebibliography}{99}
\bibitem {note1}Our computed $\mathcal{D}_{3}$ value \cite{Bidinosti03} has
been confirmed by a very recent precise NMR measurement with hyperpolarized
gas, reported in Ref. \cite{Hayden04}: $0.140(6)$~$\mathrm{m}^{\mathrm{2}%
}\mathrm{/s}$ for pure $^{3}\mathrm{He}$ gas at $296$~$\mathrm{K}$ and
$1.00$~$\mathrm{torr}$, in perfect agreement with $D_{\mathrm{gas}%
}=1.967\times\left(  296/300\right)  ^{1.71}\times760=0.1461$~$\mathrm{m}%
^{\mathrm{2}}\mathrm{/s}$.
\bibitem {note2}In the literature, experiments similar to ours would sometimes
be presented as `self-diffusion' measurements (here, performed in a confined
geometry), because atoms just move with thermal velocities. They are, hereby,
explicitly opposed to the so-called gas `diffusion' measurements where atoms
diffuse in response to a pressure gradient, that is applied across the porous
sample. In this very common type of measurements, an additional drift velocity
is imposed to the atoms by the pressure-driven motion.
\end{thebibliography}
\end{document}